\documentclass[11pt,preprint]{aastex}

\usepackage{lineno}
\usepackage{amsmath}
\usepackage{longtable}
\usepackage{lscape}
\usepackage{multirow}

\usepackage[T1]{fontenc}
\usepackage{textcomp}
\usepackage[variablett]{lmodern}
\usepackage{listings}

\slugcomment{Accepted by ApJ (30 January 2016)}
\shorttitle{LAT Source Classification}
\shortauthors{}

\makeatletter
\def\LT@makecaption#1#2#3{%
  \LT@mcol\LT@cols c{\hbox to\z@{\hss\parbox[t]\LTcapwidth{%
    \sbox\@tempboxa{#1{#2 }#3}%
    \ifdim\wd\@tempboxa>\hsize
      #1{#2 }#3%
    \else
      \hbox to\hsize{\hfil\box\@tempboxa\hfil}%
    \fi
    \endgraf\vskip\baselineskip}%
  \hss}}}
\makeatother

\begin{document}

\title{Classification and Ranking of Fermi LAT Gamma-ray Sources from the 3FGL Catalog using Machine Learning Techniques}

\author{P.~M.~Saz Parkinson\altaffilmark{1,}\altaffilmark{2},\altaffilmark{3},
  H.~Xu\altaffilmark{4} , P.~L.~H.~Yu\altaffilmark{4}, D.~Salvetti\altaffilmark{5},
M.~Marelli\altaffilmark{5}, A.~D.~Falcone\altaffilmark{6}}

\altaffiltext{1}{Department of Physics, The University of Hong Kong, Pokfulam Road, Hong Kong, China}
\altaffiltext{2}{Laboratory for Space Research, The University of Hong Kong, Hong Kong, China}
\altaffiltext{3}{Santa Cruz Institute for Particle Physics, University
  of California, Santa Cruz, CA 95064}
\altaffiltext{4}{Department of Statistics \& Actuarial Science, The
  University of Hong Kong, Pokfulam Road, Hong Kong, China}
\altaffiltext{5}{INAF - Istituto di Astrofisica Spaziale e Fisica Cosmica Milano, via E. Bassini 15, 20133, Milano, Italy}
\altaffiltext{6}{Department of Astronomy \& Astrophysics, The Pennsylvania State University, University Park, PA 16802}

\begin{abstract}
We apply a number of statistical and machine learning techniques to
classify and rank gamma-ray sources from the Third $Fermi$ Large Area
Telescope (LAT) Source Catalog (3FGL), according to their likelihood
of falling into the two major classes of gamma-ray emitters:
pulsars (PSR) or Active Galactic Nuclei (AGN). Using 1904 3FGL sources
that have been identified/{\it associated} with AGN (1738) and PSR (166), we train (using
70\% of our sample) and test (using 30\%) our algorithms and find that the best overall accuracy ($>$96\%) is
obtained with the Random Forest (RF) technique, while using a logistic
regression (LR) algorithm results in only marginally lower accuracy.
We apply the same techniques on a sub-sample of 142 known
gamma-ray pulsars to classify them into two major
subcategories: {\it young} (YNG) and {\it millisecond} pulsars (MSP). 
Once more, the RF algorithm has the best overall accuracy
($\sim$90\%), while a boosted LR analysis comes a close second. We apply our two best models (RF and
LR) to the entire 3FGL catalog, providing predictions on the
likely nature of {\it unassociated} sources, including the likely {\it
  type} of pulsar (YNG or MSP). We also use our predictions to shed
light on the possible nature of some gamma-ray sources with known associations (e.g. binaries,
SNR/PWN). Finally, we provide a list of plausible X-ray counterparts
for some {\it pulsar candidates}, obtained using {\it Swift}, {\it Chandra},
and {\it XMM}. The results of our study will be of interest for both
in-depth follow-up searches (e.g. pulsar) at various wavelengths, as well as for broader population studies.
\end{abstract}

\section{Introduction} \label{introduction}

The field of gamma-ray ($>$100 MeV) astronomy has long been plagued by
the double problem of low number of photons (and hence, sources), and
their poor  characterization. Indeed, the angular resolution of gamma-ray
instruments is typically measured in degrees or arcminutes at best, also being a strong
function of energy (improving with increasing energy).  Thus, one of
the earliest gamma-ray catalogs, the Second COS-B
catalog (2CG), contained only 25 sources and 21 of them were {\it
  unidentified}~\citep{COSB}, while two decades later, the Third EGRET Catalog (3EG)
contained 271 sources ~\citep{3EGcatalog}, almost two thirds of which remained {\it
  unidentified}, despite intense follow-up efforts~\citep{Thompson08}.

 The {\it Fermi} Large Area Telescope (LAT), launched in 2008, represents a giant leap in capabilities
  compared to past instruments. With its silicon-strip detector technology, wide
  field of view (2.4 sr), and high duty cycle ($>$95\%), the LAT has
  not only already detected over 1000 times more photons than EGRET did, but
  these cover a far broader energy range (20 MeV to $>$300 GeV)
  and are much better characterized~\citep{LATinstrument}.


Within a few months of launch, the LAT team released a first list of 205
highly-significant ($>10 \sigma$) sources. Known as the {\it Bright
  Source List} (also referred to 
as 0FGL~\citep{FGL0}), this list represented a big improvement over 
previous catalogs, as illustrated by the fact that fewer than 20\% of
0FGL sources were {\it unassociated}\footnote{The {\it identification} of a gamma-ray
  source typically requires a correlated timing
    signature at different wavelengths, whereas an {\it association}
    is less stringent, being based
    only on positional coincidence. The greatly reduced
    positional uncertainties of LAT sources, compared to previous
    experiments, has reduced the number of plausible associations per
    source, making it now more useful to talk in terms of {\it unassociated} sources,
  rather than {\it unidentified} ones.} at the time of
publication\footnote{At present, only $\sim$5\% of 0FGL sources remain unassociated.}. Full-fledged {\it
  Fermi}-LAT catalogs have been released periodically since then,
based on 11 months~\citep[1FGL, ][]{FGL1}, two
years~\citep[2FGL, ][]{FGL2}, and most recently, four years of
data~\citep[3FGL, ][]{FGL3}.  The number of known ($>$ 100 MeV) gamma-ray sources now stands at
over 3000, with approximately one third of these in the {\it
  unassociated} category~\citep{FGL3}. Figure~\ref{source_fraction}
shows the fraction of 3FGL sources in the three broad categories,
Active Galactic Nuclei (AGN), pulsars (PSR), and Unassociated sources as a function of the (4-year) significance of the source. The sharp drop in
the fraction of gamma-ray sources that are pulsars, with decreasing
significance (from $\sim20\%$ of $>20\sigma$ sources, to $\sim10\%$ of
$\sim10\sigma$ sources), in contrast to the fraction of AGN
(a relatively uniform $\sim60\%$) suggests that the discovery space
for new (relatively gamma-bright) pulsars remains significant.

In addition to uncovering large numbers of new gamma-ray sources, the
improved sensitivity of the LAT also enables a much better characterization
of these sources, facilitating their identification. Among the
earliest scientific results from the LAT, was the discovery of a large population of {\it radio-quiet} gamma-ray
pulsars~\citep{LATBSPs, sazparkinson10}, using a new blind-search
technique developed specifically for the very long time series
expected with the LAT~\citep{atwood06}. These findings confirmed
some predictions that many of the unidentified EGRET sources in the Galactic plane were
pulsars~\citep[e.g.][]{Yadigaroglu95}, favoring {\it outer gap} pulsar
models~\citep{Cheng86,Romani96,romani14}, where the gamma rays are generated in
the outer magnetosphere, as opposed to {\it polar cap} models, where
the emission comes from closer to the neutron star surface~\citep[e.g.][]{harding98}. 

Perhaps even more surprising than the large number of young, {\it
  radio-quiet} pulsars discovered by the LAT was the detection
of a large population of gamma-ray {\it millisecond} pulsars
(MSPs)~\citep{LATMSPs}, something which, though not entirely
unforeseen~\citep[e.g.][]{Harding05}, has exceeded all expectations. Through
the joint efforts of the LAT team and radio astronomers working at the major radio observatories around the world,
the {\it Fermi} LAT Pulsar Search Consortium (PSC) has been carrying out
extensive radio observations of both newly-discovered LAT gamma-ray pulsars, as well
as carrying out new pulsar searches in LAT {\it unassociated}
sources~\citep{PSC}. This has led to the discovery of over 70 new
pulsars to date~\citep[e.g.][]{ransom11}, the vast majority of which
are MSPs\footnote{For the latest list of LAT-detected gamma-ray pulsars, see {\tt https://confluence.slac.stanford.edu/x/5Jl6Bg}}.

Aside from pulsars, there are many classes of astrophysical objects that emit gamma
rays. In fact, the 3FGL Catalog lists around twenty different gamma-ray
source classes. By far the two largest classes are,
broadly speaking, pulsars (PSR) and Active
Galactic Nuclei (AGN), especially those of the {\it blazar}
variety~\citep{FGL3}. Indeed, AGN detected by the LAT can be further
subdivided into many different classes (e.g. flat-spectrum radio quasars
(FSRQs), BL Lacs, etc.). For extensive details, including the latest catalog
of LAT-detected AGN, see \cite{3LAC}.

It has been known for some time now that the two main classes of
gamma-ray sources (AGN and PSR) can be roughly distinguished by their timing and spectral
properties: AGN display variability on month-long time scales and
have energy spectra that break more softly than pulsars in
  the LAT energy band. Pulsars tend to be {\it non-variable}
(on long time scales) and have spectra with more {\it curvature},
breaking on both ends, and therefore poorly described by a simple power law, normally
requiring the addition of an exponential cut-off at a few GeV. This was illustrated graphically in the 1FGL catalog
by means of a Variability-Curvature plot~\citep[See Figure
8 from][]{FGL1}, showing pulsars and AGN clustering in opposite
corners. Indeed, a number of bright {\it pulsar candidates} were identified this
way~\citep[e.g.][]{J2339,J1311}, and later discovered to be pulsars~\citep{pletsch12,ray14}.

The large increase in the number of gamma-ray sources detected with the LAT, as
well as the somewhat crude and arbitrary (as well as subjective) nature of
visual inspection techniques, make it desirable
to develop an automated scheme to classify {\it candidate}
sources, according to their {\it predicted} source class. In recent years there has been an explosion of interest in {\it data
  science}, and the application of statistical techniques to all
fields, including astronomy~\citep[for a nice overview of recent developments in machine
learning and data mining techniques applied to astronomy,
see][]{DataMining}. Various groups have started applying these
techniques to astronomical data. Recently, ~\cite{WISE} applied the Random
Forest algorithm to the classification of variable stars using the
Wide-field Infrared Survey Explorer (WISE) data, achieving efficiencies of up
to $\sim$85\%. In the gamma-ray regime, a number of groups have
worked on both unsupervised learning~\citep{Lee12} and supervised
learning~\citep{SYBIL} techniques. Indeed, the {\it Fermi} LAT
Collaboration applied two different machine-learning techniques to the automatic
classification of 1FGL sources: logistic regression and
classification trees. By combining the two methods, a success rate of $\sim$80\% was estimated for the correct
classification of the gamma-ray source class~\citep{STATSFGL1}. An
artificial neural network approach has also been implemented and
applied to the 2FGL catalog with some promising preliminary results~\citep{SalvettiThesis}.

We have explored a large number of statistical techniques for ranking
data~\citep[e.g.][]{alvoyu} and have applied some of the most commonly used
algorithms to the problem of classification of LAT  gamma-ray sources
in the most recent 3FGL Catalog~\citep{FGL3}. Our goal is not to
  firmly establish the class of each of the $\sim$1000
  unassociated gamma-ray sources in 3FGL; rather, it is to provide an
  objective measure that quantifies the likelihood of each source of
  belonging to one of the two major classes (pulsar or AGN), for the purpose of aiding in the necessary
  follow-up studies and searches (mostly in other wavelengths) that
  {\it can} conclusively determine the nature of each individual
  source. In pointing out sources that are unlikely to belong to
  either of the big source classes, our study also serves to highlight
  gamma-ray sources that might have a more {\it exotic} origin (e.g. dark matter
  annihilation). Finally, our results may be useful for population studies,
  or to estimate the number of new gamma-ray sources in each class
  that we might expect to identify in the future.

The structure of the paper is as follows. In Section~\ref{dataselection} we describe the data sets used in
the paper as well as some of the key predictor parameters
employed. Section~\ref{algorithms} discusses the various algorithms we
considered, with an emphasis on those that proved most
successful (Random Forest and Logistic Regression).  Next, in
Section~\ref{application} , we discuss
the application of these algorithms to the 3FGL catalog and provide an
overview of our results. Finally, Section~\ref{discussion} provides a
discussion of our key results and conclusions, including our
predictions on the nature of both the {\it unassociated} sources, as well as certain
gamma-ray sources with known associations (e.g. gamma-ray binaries and
SNR/PWNe). We also provide a list of plausible X-ray counterparts for
some of our best {\it pulsar candidates}, obtained through our
follow-up program of LAT gamma-ray sources with {\it Swift}, {\it
  Chandra}, and {\it XMM}.

\section{Data and Feature Selection} \label{dataselection}


The {\it Fermi} LAT Third Source Catalog (3FGL) was publicly released
through the Fermi Science Suppport Center (FSSC\footnote{{\tt http://fermi.gsfc.nasa.gov/ssc/}}) in January
2015, with a few minor updates being posted around the time of official publication~\citep{FGL3}. The results presented here make use
of the uptdated version released on 2015 May 18 (FITS file gll\_psc\_v16.fit\footnote{The 3FGL catalog, along with other LAT catalogs, are also available as an R package, {\tt fermicatsR},
  from the Comprehensive R Archive Network webpage {\tt
    http://cran.r-project.org/web/packages/fermicatsR/}}). The 3FGL
catalog contains 3,034\footnote{Note that the various components
  associated with the Crab nebula are counted as different sources.} gamma-ray sources, of which 1010 are {\it unassociated}.

\subsection{Training and Testing Sets \label{sampling}}

Since we are interested in applying a number of {\it
  supervised} learning techniques to classify sources according to
their likelihood of falling into two broad classes (PSR and AGN), our first step involves selecting sources from
3FGL that are known to fall into these two categories.  We select all
sources that are identified or associated with pulsars (i.e. sources
classified as PSR or psr, respectively, in 3FGL), and all sources that are identified or associated
with any type of AGN, including FSRQs, BL Lacs, etc  (i.e. sources of the
following class: FSRQ, fsrq, BLL, bll, BCU, bcu, RDG, rdg, NLSY1, nlsy1, agn, ssrq, and sey\footnote{For a definition of all
these acronyms, see Table 6 of \cite{FGL3}.}). After filtering
out sources with missing values (six AGN and one
pulsar, namely PSR J1513--5908), we are left with a total of 1904
sources, of which 1738 are AGN and 166 are PSR. We created training and testing sets by randomly
selecting 70\% and 30\% of these sources, respectively. In summary, our
training sample contains 1,217 AGN and 116 PSR, while our testing sample
contains 521 AGN and 50 PSR.

Because we are also interested in the classification of pulsars into
`young' (YNG) and millisecond (MSP), we further split the known
gamma-ray pulsars into these two sub-samples. We make use of the public list of
LAT-detected gamma-ray pulsars\footnote{{\tt
    https://confluence.slac.stanford.edu/display/GLAMCOG/Public+List+of+LAT-Detected+Gamma-Ray+Pulsars}}
and cross-correlate these with the 3FGL catalog, obtaining a list of 142
known gamma-ray pulsars (77 YNG and 65 MSP) which, we randomly split up
into a training set containing 70\% of the sample (52 YNG and 47 MSP) and a
testing set with the remaining 30\% (25 YNG and 18 MSP). 

\subsection{Feature Selection}

The 3FGL Catalog contains a large number of measured parameters on each source,
covering everything from the source positions and uncertainties, to
fluxes in various bands, etc. In addition to all the parameters
included in 3FGL,  we also defined the following {\it hardness
  ratios}, following \cite{STATSFGL1},  as:

$hr_{ij} = (EnergyFlux_j - EnergyFlux_i)/(EnergyFlux_j + EnergyFlux_i)$

where $i$ and $j$ are indices corresponding to the five different LAT energy
bands defined in the 3FGL catalog: i=1: 100--300 MeV, i=2: 300 MeV --
1 GeV, i=3: 1--3 GeV, i=4: 3--10 GeV, and i=5: 10--100 GeV, respectively. The energy flux in each band is computed
by integrating the photon flux, using the measured spectral index.  

We started out with a large number (35) of potential parameters on which
to train our sample. Using a two sample t-test, we then determined how the
various parameters were correlated to each other (e.g. Signif\_Avg is
highly correlated with Flux\_Density) and removed highly correlated
($\mid$$\rho$$\mid >$0.7) parameters as well as
parameters related to the position of the source (e.g. GLAT, GLON). We then applied a log
transformation to some parameters that displayed highly skewed
distributions (e.g. Variability\_Index). We dropped sources with missing values for any of our predictor parameters, leaving
a total of 3021 sources, out of the original 3034. Table~\ref{variables} shows the 9 predictor parameters
that we ended up using in our various models for classifying PSR vs AGN, along with the range and
median of these parameters in our data sets. These include the
well-known {\it curvature} and {\it variability} parameters
(Signif\_Curve and Variability\_Index, respectively), the spectral
index and flux density of the source, the uncertainty in the energy
flux above 100 MeV (Unc\_Energy\_Flux100), and the four hardness
ratios constructed using the five different energy bands and equation described
above (i.e. $hr_{12}$, $hr_{23}$, $hr_{34}$, and $hr_{45}$). For the
MSP vs YNG classification, we added GLAT (Galactic Latitude) as a
possible predictor parameter. In Appendix~\ref{getdata} we provide the R
script used to obtain and {\it clean} the data, as well as to perform
our detailed feature selection. 

\section{Classification Algorithms} \label{algorithms}

We considered a long list of algorithms (see Tables \ref{AGNvPSR} and
\ref{YNGvMSP} for a complete list), including Decision Trees,
Support Vector Machines (SVM), a simple logistic regression (LR) model (with both
forward and backward stepwise elimination), various modified versions
of LR (e.g. Boosted LR, logistic decision trees), Random Forest (RF), as well as some combination of methods (e.g. a 2-step method
involving decision trees followed by LR). We used the {\tt RWeka} package~\citep{RWeka,Weka}, along with the
{\tt pROC} R package~\citep{pROC} to draw Receiver Operating
Characteristic (ROC) curves, commonly used in the data mining
and machine learning communities to measure the performance of
a classifier. In such a curve, one plots sensitivity (true positive rate) vs
specificity (true negative rate) for varying thresholds, allowing us
to evaluate the tradeoffs involved in each choice, with the ultimate goal being to classify
correctly the largest proportion of pulsars in our sample, while keeping the proportion of
mis-classifications as low as possible. Figures~\ref{lr_roc} and
\ref{rf_roc} show the ROC curves for our two best algorithms in the
AGN vs PSR classification, while Figures~\ref{blr_pulsars_roc} and
\ref{rf_pulsars_roc} show the corresponding ROC curves for the best
two algorithms in the YNG vs MSP classification. We define the best threshold
as that which maximizes the sum of both terms (sensitivity + specificity, see Appendix~\ref{roc} for the relevant R
scripts). We also used the\texttt{ randomForest }\citep{RFR} R package to fit random
forests and the \texttt{e1071}~\citep{e1071} R package to fit
SVM models. Ultimately, we settled on the RF algorithm, for its overall
accuracy, and LR for its slightly better sensitivity to pulsar
classification. In the following sections we provide some details on the most successful models we decided to apply to the
3FGL Catalog.

\subsection{Random Forest}\label{RFsection}

Random forest~\citep[][hereafter RF]{RF} is an ensemble learning method which uses decision
trees as building blocks for classification, regression and other
tasks. By aggregating the predictions based on a large number of
decision trees, RF generally improves the overall predictive
performance while reducing the natural tendency of standard decision trees to
over-fit the training set. RF is essentially an extension of a Bootstrap aggregating method called
Tree bagging~\citep{breiman1996bagging}. First of all, tree bagging for classification
generates $B$ different training sets by sampling with replacement
from the original training set. Then it builds a separate tree for
each training set, resulting in $B$ fitted trees, and finally for each
new observation, it generates the predicted class probability by
taking the average of the $B$ predicted class probabilities from all $B$ fitted trees.


RF provide an improvement over bagged trees by way of a
random small tweak that decorrelates the trees. Its main idea is that
in each step of identifying the best split of a node in the tree growing stage, a
random sample of $m$ parameters drawn from all the predictor parameters
is used for consideration in selecting the best split for that node.
Typically, a value of $m\sim\sqrt{p}$ is used, where $p$ is the total number of
parameters tried at each split. By forcing each split to consider only a subset of the predictor
parameters, the splits will not always be constructed from the
strongest predictor but some other potential strong predictors,
thereby making the RF more reliable.  It is thus advantageous to use RF when there exists multicollinearity
in the predictor parameters. The detailed algorithm can be found in
the book by \citet{james2013introduction}.

In this paper, we make use of the \texttt{randomForest} package \citep{RFR} in R. In order to ensure the stability of our
results, we grew a large number of trees (10,000 vs the default of
500) and scanned over a range of values of $m$. We found that a value of
$m$=2 was optimum, giving an OOB\footnote{``Out-of-bag'', in the sense
  that it is based on the portion of the data not already used for training the
  original tree, thus providing an internal estimate of the error
  which is expected to be comparable to that obtained with a final
  independent {\it testing} data set. See \cite{RF} for details.} estimate of the error of 2.3\%.

The \texttt{randomForest} package can also generate a variable importance measure for each parameter in terms of the mean decrease
in accuracy (MDA), which is computed from permuting the OOB data: For each tree,
the prediction error on the OOB portion of the data is recorded
(error rate in the classification). Then the same is done after permuting each predictor variable. The difference between
the two are then averaged over all trees~\citep{RFR}.


The \texttt{randomForest} package
also computes the {\it proximity} measure, which, for each pair of
elements (i, j), represents the fraction of trees in which elements i
and j fall in the same terminal node. This can be used to calculate
the ``outlyingness'' of a source, as the reciprocal of the sum of
squared proximities between that source and all other sources in the
same class, normalized by subtracting the median and dividing by the
median absolute deviation, within each
class~\citep{RFR}. Figure~\ref{outlyingness} shows the {\it outlyingness} of all 3FGL
sources, with respect to the PSR and AGN classes. Note that most
pulsars have large values of ``AGN outlyingness'', while most AGN have
large values of ``PSR outlyingness''. A large value of ``outlyingness'' along both
  axes could imply a different gamma-ray source class altogether
  (i.e. non-Pulsar {\it and} non-AGN). 


\subsection{Logistic Regression}\label{LRsection}

Logistic regression (hereafter LR) is a very popular probability model that was developed
by \citet{cox1958} and \citet{walker1967}. The model can be
used to predict a binary response based on one or more predictor parameters
(features). For an 0/1 binary response variable $y$ and $p$ predictor
parameters $x_{1},...,x_{p}$, the logistic regression model can be
written as
\[
y\mid x\sim Bernoulli(\pi)
\]
and
\[
logit(\pi) \equiv \log\frac{\pi}{1-\pi}=\beta_{0}+\beta_{1}x_{1}+...+\beta_{p}x_{p}
\]
where $\beta=(\beta_{0},\beta_{1},...,\beta_{p})$ is a vector of
unknown parameters which can be estimated by the maximum likelihood
method. For further details on logistic regression, including its use
as a classifier, see \cite{LR}.

Both forward and backward stepwise methods are considered in parameter
selection. Forward stepwise method starts with no predictor parameters
in the model, and then recursively adds parameters one at a time
according to the Akaike Information Criterion (AIC).  
Backward stepwise method is performed similarly except that it starts
with all predictor parameters in the model and proceeds by dropping
parameters one at a time. In our study, we found that both methods gave the same result.

\subsection{Boosted Logistic Regression}\label{BLRsection}

Boosted logistic regression is a powerful classifier based on additive
logistic regression fitted by stage-wise optimization of the Bernoulli
log-likelihood \citep{friedman2000additive}. In a two-class problem,
the additive logistic regression is of the form:
\[
\log\frac{P(y=1\mid x)}{P(y=0\mid x)}=\sum_{m=1}^{M}f_{m}(x)
\]
where $f_{m}(x)$ is learned in fitting the $m$th logistic regression
on a weighted training data. Its basic idea is to adaptively change
the frequency weights of the observations in the training data according
to the performance of the previous logistic model. Unlike fitting
a single model to the data which may not fit the data well or potentially
suffer from over-fitting, the boosting approach instead, by sequentially
fitting the logistic regression models, gradually improves the model
fit in the regions where it originally did not perform well.

\citet{friedman2000additive} developed an algorithm for fitting additive
logistic regression models, named as the LogitBoost algorithm which
can be implemented using the LogitBoost function in the RWeka package
\citep{RWeka,Weka}.

\subsection{Model Building Procedure}

In order to determine the optimal cutoff probability value of the
prediction generated from each model, we adopted a 10-fold
cross-validation method. The procedure of model building is summarized as follows:

\begin{enumerate}
\item We partitioned our 3FGL data set at random into 70\% for
  training and 30\% for testing, as described in Section~\ref{sampling}.
\item We divided the training set randomly into 10 equal-size subsets.
\item We used 9 subsets to build a model and apply the fitted model to
  test on the remaining subset. We then repeated this procedure for
  all 10 subsets until all the subsets were tested.
\item We obtain the ROC curve based on all the tested subsets and determine the best cutoff value.
\item Using the training data to build a model, we then apply the fitted model and best cutoff value to generate our prediction (``PSR" or ``AGN") for the testing data.
\end{enumerate}

\section{Application of the algorithms to 3FGL} \label{application}

Figure~\ref{mda1} shows the relative importance of the input
parameters in the RF model, expressed as the mean
decrease in accuracy, as described in Section~\ref{RFsection}. Note that in some cases we
have transformed the parameters by taking the log, due to the skewness
of the distributions. We note that, perhaps not surprisingly, the
curvature significance and variability parameters (Signif\_Curve and
Variability\_Index) are two of the three most important predictor
variables, while the uncertainty in the energy flux
(Unc\_Energy\_Flux100) also turns out to
be very important, perhaps as a proxy for the quality of the spectral
fit of the source. In the case of the classification of pulsars
into YNG and MSP, we added GLAT (Galactic Latitude),
which we found to be useful in discriminating between the two classes
(In fact, it is the second most important parameter, see
Figure~\ref{mda2}). Table~\ref{logit}, on the other hand, shows the values of the
parameters ($\beta_{0}, \beta_{1},...$) corresponding to the
various predictor parameters in the best logistic regression model
(backwards stepwise), giving also an indication of the significance of
each one (Variability\_Index being the most significant, in this case).

\subsection{Results}

After applying a large number of algorithms to the problem of
gamma-ray source classification, we concluded that the RF technique 
provides the overall best accuracy\footnote{Defined as the
  number of correctly classified AGN and PSR, divided by the total
  testing sample size.} (96.7\%), while LR (with backward stepwise elimination)
provides only slightly lower overall accuracy (94.7\%), but a better
sensitivity to pulsar identifications (98\%, vs 96\% for RF). Table~\ref{AGNvPSR} provides a summary the results of all the
various algorithms we tried, as applied to the problem of classifying AGN and
PSR. 

The performance of the models, given in Tables~\ref{AGNvPSR} and
  \ref{YNGvMSP}, is clustered into two groups: the best algorithms,
which perform basically the same, and the worst algorithms, which are 
also basically the same. The best algorithms are all those tree-type
models such as Decision Tree and RF (CV) while the worst ones are
those linear-type models such as LR and Boosted LR (CV). Therefore,
tree-type models seem better than linear-type models in classifying 
AGN and PSR. From Table~\ref{YNGvMSP}, the best performing algorithms
in the case of YNG vs MSP are Boosted LR and RF, which are the only two
ensemble methods considered here.  Ensemble methods aim at combining
many models to form the final classification (e.g., RF combines many
tree models while boosted LR combines many LR models). Therefore, we
believe that combining models can help improve the classification of
young vs millisecond pulsars. The small scatter in the performance of
the algorithms is likely due to the imbalanced nature of the data
sets. Indeed, we see that even the worst performing algorithm has an overall accuracy
of $>$93\%. It is worth considering an additional performance measure
called F1 score (or F-measure), which is defined as the harmonic mean of precision
and recall, therefore conveying a balance between these two
quantities~\citep{F1}. This F1 score for PSR ranges from 0 to 1, with
a larger value implying a better performance in classifying PSR. The
F1 scores have a wider range (see last column in Table~\ref{AGNvPSR}),
and the four tree-type models perform better than the linear-type models for the classification
of PSR and AGN, with the RF still returning the best performance.

To test the robustness of our results, we ran both the RF and LR
algorithms ten times, randomly selecting different training and
testing sets in each case and found consistent results. Both
methods returned an overall accuracy of $\sim$96\%, with the RF
technique performing marginally better than LR.

For our analysis of the pulsar population, using a much smaller sample of 142 known
gamma-ray pulsars (77 YNG and 65 MSP), we again found that the RF 
algorithm returns the best overall accuracy (90.7\%), while a boosted
logistic regression analysis comes a close second (88.4\%). We
caution, however, that these results are based on a relatively small testing
sample of only 43 pulsars (25 YNG and 18 MSP). Table~\ref{YNGvMSP}
summarizes the results obtained by the various algorithms in the
classification of young (YNG) vs millisecond pulsars (MSP). 

Having settled on the best models, we then applied these to the
entire 3FGL catalog (that is, all 3,021 sources for which predictor
parameters are available and for which our models could therefore be
  applied). Table~\ref{allresults} shows a portion of these results
  (the full table being available electronically from the
  journal\footnote{Also at {\tt
      http://www.physics.hku.hk/$\sim$pablo/pulsarness/Step\_08\_Results.html}}
  \footnote{Also at {\tt
      http://scipp.ucsc.edu/$\sim$pablo/pulsarness/Step\_08\_Results.html}}). In
  the next section we go over some of the implications and follow-up
multi-wavelength studies based on these results.

\section{Discussion and Conclusions} \label{discussion}

\begin{figure}
\epsscale{1.0}
\includegraphics[width=6in]{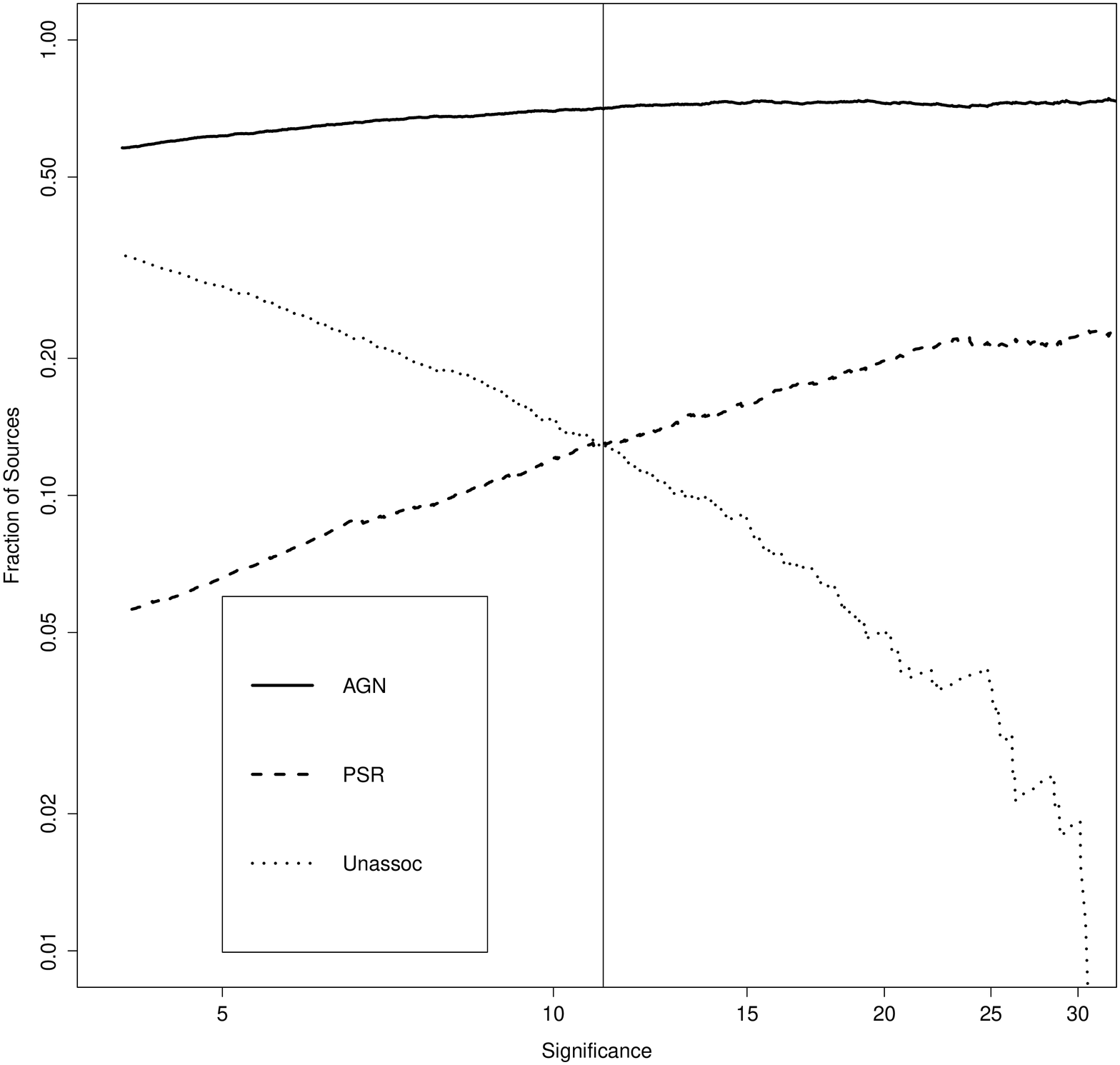}
\caption{Fraction of 3FGL sources in the three major categories (AGN, PSR, and Unassociated) as a
function of the (4-year) significance of the source. The vertical line represents the lowest significance of a
pulsar found in a blind search of gamma-ray data so far (PSR J1023-5746, 11.1$\sigma$).
 \label{source_fraction}}
\end{figure}

\begin{figure}
\epsscale{1.0}
\includegraphics[width=6in]{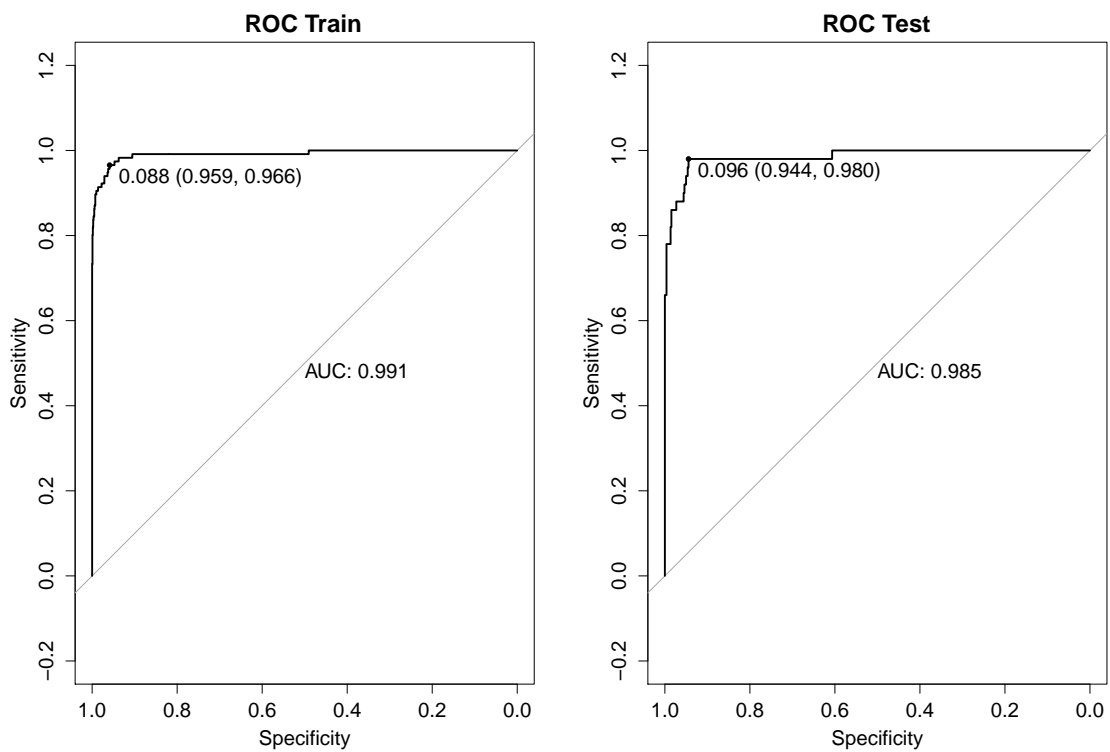}
\caption{ROC curves for the LR model for the AGN vs PSR classification. On the {\bf left} is the curve
  obtained on the training set, showing that the best threshold
  (P=0.088) results in a specificity of 0.959 and sensitivity of
  0.966. The one on the {\bf right} shows the corresponding values for the
  model applied to the testing set.
 \label{lr_roc}}
\end{figure}

\begin{figure}
\epsscale{1.0}
\includegraphics[width=6in]{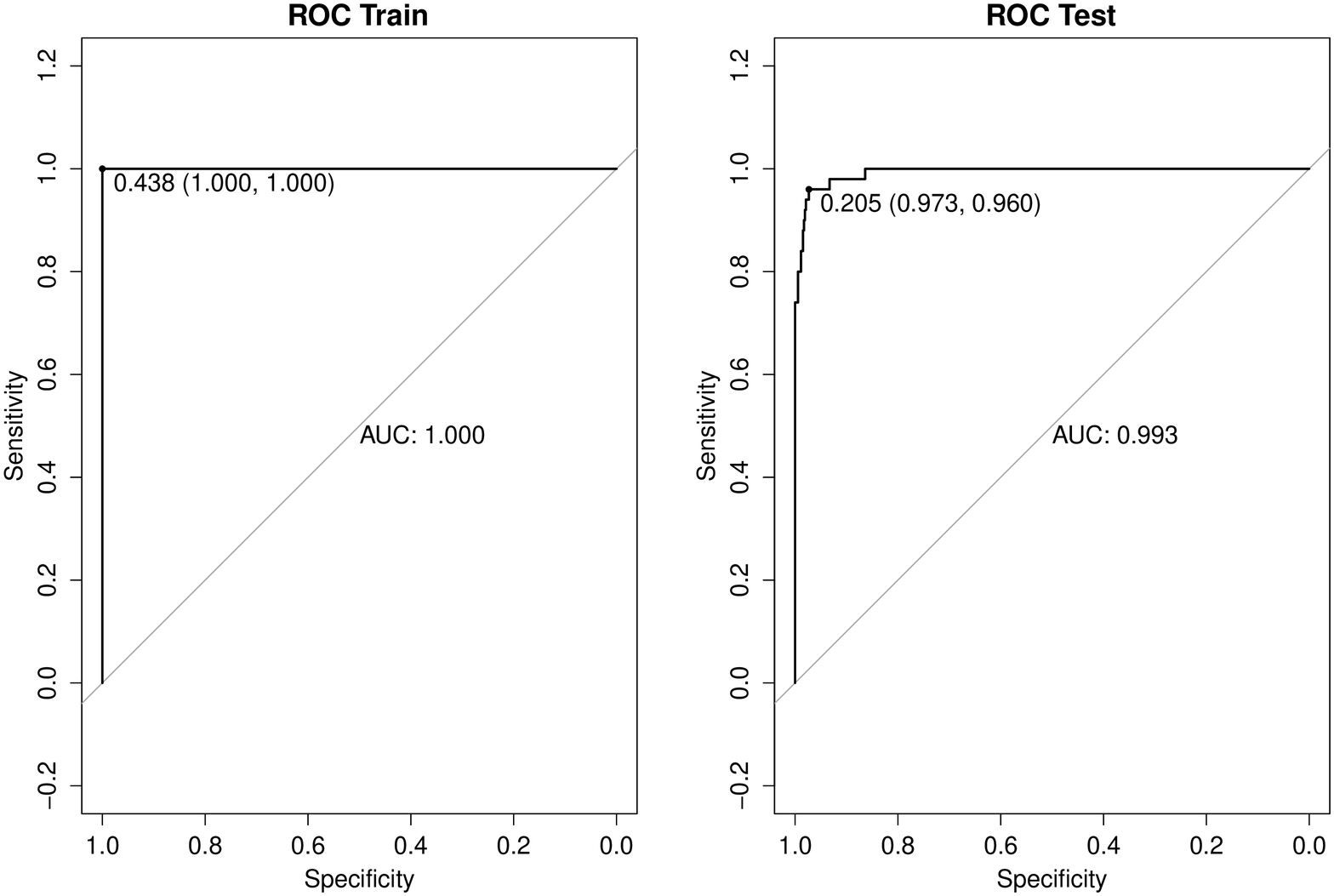}
\caption{ROC curves for the RF model for the AGN vs PSR classification. On the {\bf left} is the curve
  obtained on the training set, showing that the best threshold
  (P=0.438) results in a specificity of 1.0 and sensitivity of
  1.0. The plot on the {\bf right} shows that the corresponding values for the
  model applied to the testing set results in 0.973 and 0.960, respectively.
 \label{rf_roc}}
\end{figure}

\begin{figure}
\epsscale{1.0}
\includegraphics[width=6in]{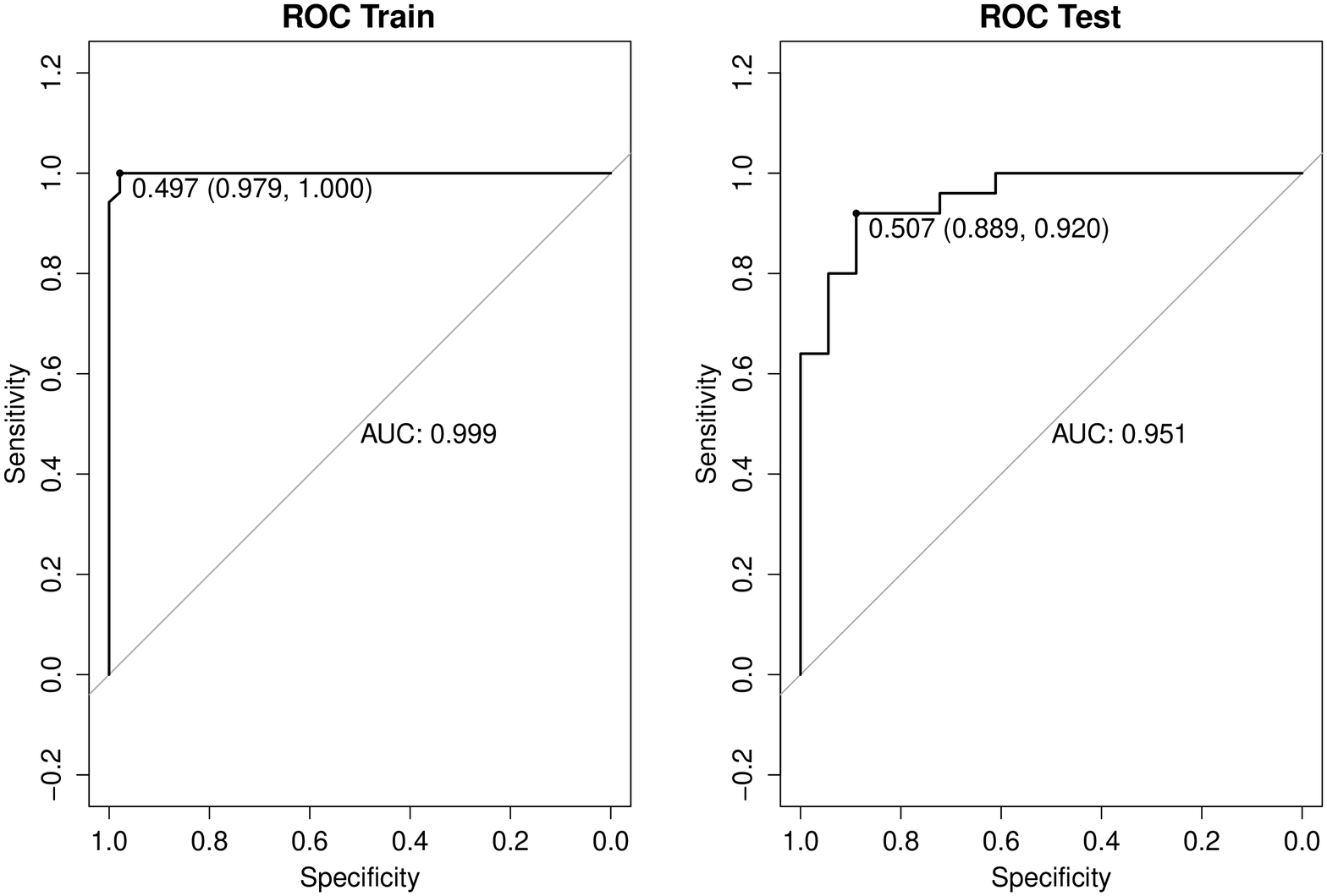}
\caption{ROC curves for the Boosted LR model for the YNG vs MSP classification. On the {\bf left} is the curve
  obtained on the training set, showing that the best threshold
  (P=0.497) results in a specificity of 0.979 and sensitivity of
  1.0. The plot on the {\bf right} shows that the corresponding values for the
  model applied to the testing set results in 0.889 and 0.920, respectively.
 \label{blr_pulsars_roc}}
\end{figure}

\begin{figure}
\epsscale{1.0}
\includegraphics[width=6in]{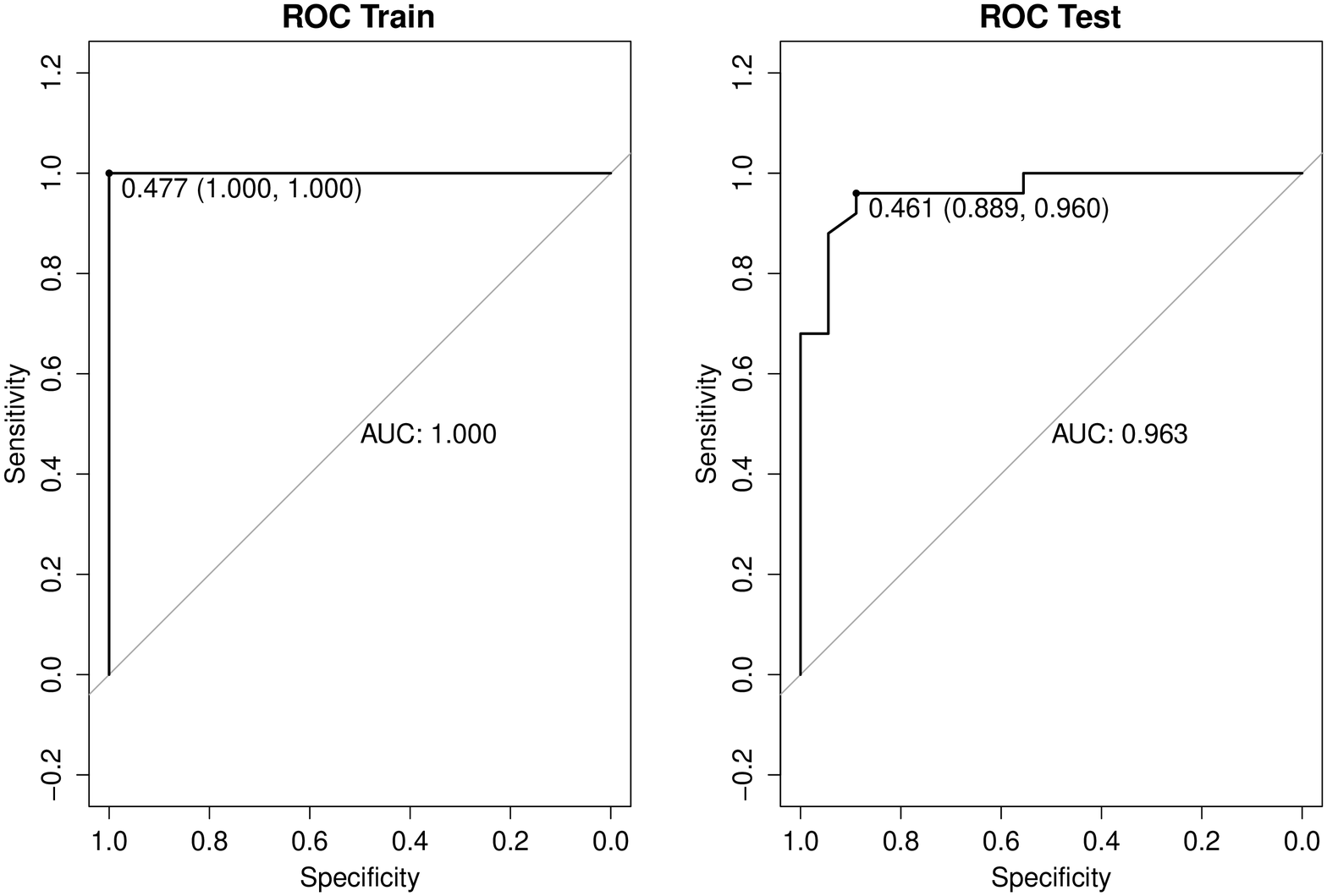}
\caption{ROC curves for the RF model for the YNG vs MSP classification. On the {\bf left} is the curve
  obtained on the training set, showing that the best threshold
  (P=0.477) results in a specificity of 1.0 and sensitivity of 1.0. The one on the {\bf right} shows the corresponding values for the
  model applied to the testing set results in 0.889 and 0.960 respectively.
 \label{rf_pulsars_roc}}
\end{figure}

One of the main goals of our investigations is to identify the most
promising {\it unassociated} gamma-ray sources to target in pulsar
searches, both in blind gamma-ray searches and in radio
searches. As discussed in Section~\ref{introduction}, these searches have been very
fruitful in the past, and as indicated by Figure~\ref{source_fraction}, the discovery potential remains
significant. 

Overall, we find that of the 1008 {\it unassociated} sources for which we have a
prediction, in 893 cases the RF and LR algorithms are in agreement
(334 being classified as likely PSR and 559 as likely AGN). Out of the 334 unassociated sources
classified as likely PSR, 309 resulted in a consistent sub-classification using
both the RF and Boosted LR algorithms (194 of these being classified
as likely YNG pulsars, while 115 as likely MSPs). In Table~\ref{unassoc} we provide a list of
the most significant ($>10\sigma$) 3FGL unassociated sources which our methods (both
RF and LR) predict to be pulsars. While a $10\sigma$ cutoff is
somewhat arbitrary (and since we provide predictions for all sources,
searchers are free to set their own thresholds), we should keep in mind that no pulsars have been
found in blind searches of gamma-ray data below this significance (see
Figure~\ref{source_fraction}), so it is probably safe to say that most
pulsars found in the future will also be above this cutoff. Indeed, if
we consider only sources above 11$\sigma$ (roughly the lowest
significance for a pulsar found in a gamma-ray blind search), we note
that there are $\sim$1000 sources, of which only $\sim$125 are unassociated. Our
algorithms predict that roughly 75\% of these should be pulsars, of
which two thirds are predicted to be YNG and the remaining third
MSP. As an indication of how realistic these numbers are, we
point out that the discovery of an additional $\sim$90-95 LAT pulsars
would bring the percentage of pulsars within these $>11\sigma$ sources
up to $\sim$22\%, or roughly the same percentage found among the 239
most significant ($>32.9\sigma$) 3FGL sources, 100\% of which have
known associations. Of course, the well known lack of correlation
between radio and gamma-ray fluxes of pulsars also means that radio searches of less significant LAT
sources (led by the PSC) will continue to produce new gamma-ray pulsar
discoveries. It is likely that these will mostly continue to be in the
MSP category, since a significant fraction ($\sim$50\%) of the young 
gamma-ray pulsars that are below threshold for LAT blind search discoveries will likely turn out to be radio-quiet, while
a large fraction of young pulsars that lie along the Galactic plane have probably
already been discovered in existing deep radio surveys.

We note that some of our predictions have already been confirmed by
the latest pulsar searches. Recently, for example, the young
pulsar J1906+0722 was discovered in a gamma-ray blind search of 
3FGL J1906.6+0720~\citep{clark15} while the millisecond
pulsar (MSP) J1946-5403 was discovered in a targeted radio search of
3FGL J1946.4-5403~\citep{camilo15}. In some cases, athough pulsations
have not yet been discovered, the presence of a millisecond pulsar is strongly
suggested by multiwavelength observations
(e.g. 3FGL~J1653.6-0158~\citep{romani14}, 3FGL~J0523.3-2528~\citep{strader14,xing14},
3FGL~J1544.6-1125~\citep{bogdanov15},
3FGL~J2039.6-5618~\citep{salvetti15,romani15}), in
agreement with our predictions (see Table~\ref{unassoc}).

Searches for new gamma-ray pulsars need not be limited to {\it
  unassociated} sources. Indeed, other known gamma-ray source
classes like supernova remnants (SNRs) and pulsar wind nebulae (PWNe)
are {\it known} to be related to pulsars, and it is often hard to
disentangle the emission coming from the pulsar from that of the
remnant or PWN. Thus, it is worth looking more closely into those
sources that have been classified as SNR/PWNe, in the hope that a new
(as yet undiscovered) pulsar could be found in its
midst. Table~\ref{snrpwn} provides our model (LR and RF) predictions
for 3FGL sources with claimed SNR or PWN associations\footnote{We
  include 3FGLJ1119.1-6127  and 3FGLJ1124.5-5915, even though these sources are formally associated with
  PSRs J1119-6127 and J1124-5916 in the 3FGL catalog, rather than with
  their respective SNRs.}, which includes most of the likely GeV SNRs
(27/30) and about half of the marginal ones (8/14), as reported
in the ``First Fermi LAT Supernova Remnant
Catalog''~\citep{SNRCat}. Of the 9 SNRs from the LAT SNR catalog not
listed here (3 firm and 6 marginal), one corresponds to a LAT source
dropped from our analysis, as described in Section~\ref{dataselection}
(3FGLJ2021.0+4031e, associated with Gamma Cygni), while the remaining
8 have no corresponding 3FGL source. We find that a significant number of these sources are classified by both the RF and
LR algorithms as likely pulsars, in some cases with very large
probability. Indeed, as many as 14 of the likely SNRs from the LAT SNR
catalog have a P$>$0.95 in the LR algorithm, scoring highly in the RF
algorithm too, including such famous SNRs as IC443, Cas A, or the
Cygnus Loop. For more details on other potential associations we recommend 
consulting the ``Census of high-energy observations of Galactic
supernova remnants\footnote{{\tt http://www.physics.umanitoba.ca/snr/SNRcat/}}''\citep{Ferrand12}.
We should add that \cite{SNRCat} reported an upper limit
of 22\% on the number of GeV candidates falsely identified as SNRs, so finding 6--7 new
gamma-ray pulsars among these sources would still be consistent with
the LAT SNR Catalog results.

Gamma-ray binaries are another class of gamma-ray source that has been
predicted to be associated with pulsars~\citep{Dubus06,Dubus13}. Looking at the results of our models
as applied to the four LAT-detected gamma-ray binaries (Table~\ref{binaries}), we see that
all of them are, indeed, predicted to be pulsars (specifically, of the
YNG variety). It may, therefore, be worth considering our YNG {\it
  pulsar candidates} as also being {\it gamma-ray binary candidates},
especially as it may be simpler to discover a gamma-ray binary orbital
modulation than to discover pulsed emission from such systems.

We also considered sources with large values of {\it outlyingness},
as discussed in Section~\ref{RFsection}. Table~\ref{out} lists those
3FGL sources with large ($>$75) values of PSR and AGN
outlyingness. We looked into the five sources highlighted by \citet{SYBIL} as being
the top``outliers'', among the high-latitude 2FGL sources. Out of the
five, four are classified by our algorithms clearly as MSPs (indeed,
one of them, PSR~J0533+6759 has already been discovered), while the
remaining source (now known as 3FGL J1709.5--0335) is classified by
both our RF and LR algorithms as an AGN.

Finally, looking at the results of our predictions as applied to the set of 1904 3FGL sources associated with AGN
or PSR (i.e. our combined training and testing set), it is worth
considering how {\it consistent} the two algorithms are with each
other, in addition to how accurate they are. We find that RF and LR are in agreement in $>$95\% of cases (1825 sources). Of all of these,
we find only 13 sources where our predicted class differs from that
given in the 3FGL catalog. While it is perfectly natural to expect all
of these associations to be correct, given the small number, we
provide a list of these sources in Table~\ref{misclassifications}, in
case some may deserve further investigation. We note that in all but one case, this
misclassification involves a source that in 3FGL has been associated
with an AGN, while our algorithms predict a PSR (in most cases
of the MSP variety). In the case of PSR J1137+7528 (clearly identified in the LAT by its pulsations),
the reason for the bad model prediction is likely due to the poor
  spectral fit arising from the low source significance (4.3$\sigma$). Indeed, two out of the 5 spectral
energy bins in 3FGL are only upper limits, and the resulting power-law
fit provides a perfectly acceptable model for the spectrum, as is
usually the case with AGN.

\subsection{X-ray observations}

As discussed in Section~\ref{introduction}, our goal in applying
machine learning techniques to the entire 3FGL catalog was not so much
to establish conclusively the class of individual sources, but rather
to identify the most promising sources for further investigation.

Uncovering the nature of gamma-ray sources usually requires a coordinated
multi-wavelength effort with many instruments. X-ray observations can be particularly
useful in blind searches for gamma-ray pulsars, given the much better
angular resolution of X-ray instruments and the sensitivity of pulsar
observations to uncertainties in the position being searched~\citep{dormody11,sazparkinson14}. Furthermore, X-ray
observations of pulsars are also beginning to shed light on possible
differences between {\it radio-loud} and {\it radio-quiet}
pulsars~\citep{Marelli15}. 

In this section we describe our efforts to use X-ray observations
in the search for new gamma-ray pulsars among some of the most
promising unassociated LAT sources. Over the past several years, we
have observed a number of bright LAT unassociated gamma-ray sources with {\it Chandra} and {\it XMM}, currently the most sensitive
instruments in the $\sim$1--10 keV band~\citep[e.g.][]{sazparkinson14}. It is beyond the scope of this paper to
carry out an exhaustive analysis of all the X-ray observations of LAT
sources, many of which have, in any case, been published and discussed
elsewhere~\citep[e.g.][]{cheung12}. Here, we briefly discuss six interesting LAT sources for
which we (PI: Saz Parkinson) obtained either {\it Chandra} or {\it
  XMM} observations. 

We performed a standard reprocess, analysis, and source detection in
the 0.3-10 keV energy band of the {\it XMM-Newton} and 
{\it Chandra} observations, following~\cite{Marelli15}. For each of
the X-ray sources inside the gamma-ray error ellipse, we
performed a spectral analysis. After extracting the spectra, response
matrices, and effective area files, we fitted a power-law model using
either the $\chi^2$ statistic of the C-statistic~\citep{Cash79} in
case of a negligible background (the case of {\it Chandra} sources).
Unfortunately, the low statistics in some cases prevented us from an accurate
spectral characterization. For sources with a low number of counts
(typically fewer than $\sim$ 30), we fixed the column density to the
value of the integrated Galactic N$_H$~\citep{Kalberla05} and, if
necessary, the photon index to 2. We computed the gamma-ray to
X-ray flux ratio. As reported in \cite{Marelli11,Marelli15}, this
could give important information on the nature of the source. Finally,
we computed the predicted 5$\sigma$ upper limit on a detection, based
on the signal-to-noise. The detailed results of our X-ray analyses are
presented in Table~\ref{counterparts2}. In the following paragraph we
summarize some of our key findings. 

3FGL J1035.7-6720 and 3FGL J1744.1-7619 were both observed for $\sim$25 ks
with XMM (obsIds 0692830201 and 0692830101), and show the
presence of possible X-ray counterparts at (RA, DEC) of (158.8652,
-67.3371) and (266.0030, -76.3205), respectively. LAT sources 
3FGL J0212.1+5320 and 3FGL J0933.9-6232, both strong MSP candidates, were
observed with {\it Chandra} for $\sim$30 ks (obsid 14814) and $\sim$45 ks (obsid 14813), respectively, and show
the nearest possible X-ray counterparts at (33.0439, 53.3607) and (143.5049,
-62.5077), within the LAT error ellipse (e.g. see Figure~\ref{J0212}). Source 3FGL J1214.0-6236, coincident with
SNR G298.6+0.0 was observed with {\it Chandra} for $\sim$20 ks (obsid
14889) and shows a potential counterpart at (183.4307, -62.5936), as shown in Figure~\ref{J1214}. 3FGL J1405.4-6119, coincident with
SNR G311.5+0.3, was observed with {\it Chandra} for 13 ks (obsid
14888), showing a possible counterpart at (211.3103, -61.3077). 

In addition to the dedicated {\it Chandra} and {\it XMM} observations,
many other LAT sources have been observed with less sensitive
instruments. Indeed, since the launch of {\it Fermi}, we have been
using the {\it Swift} X-ray telescope~\citep{gehrels04} to carry out follow-up observations of LAT unassociated
sources (\cite{stroh13}, Falcone et al.  (2016), {\it in preparation}), as part of an ongoing multi-year {\it
  Fermi} Guest Investigator program (PI: Falcone). This has resulted,
to date (through 2015-12-21), in the detection of almost 1900 sources 
(with SNR $>$ 3)\footnote{{\tt http://www.swift.psu.edu/unassociated/}}. We selected only those X-ray sources within a radius of 1.2 times the
semimajor radius of the 95\% confidence region of our best pulsar
candidates (those predicted to be pulsars by both the RF and LR
methods). Table~\ref{counterparts} provides a list of these $\sim$90
potential counterparts. For each potential counterpart we convert the
measured count rate into an estimated flux in the 0.1--2.4 keV band by assuming a
power law spectrum of index -2 and estimating the Galactic Hydrogen
column density using the method of \cite{Willingale13}. Given the known
faintness of pulsars in the X-ray band~\citep[e.g.][]{Marelli11}, our flux estimates can be used
to likely rule out any bright {\it Swift} source (e.g. with flux greater than
$\sim5\times10^{-13}$ erg cm$^{-2}$ s$^{-1}$) as a plausible X-ray
counterpart of a pulsar \footnote{On this point, see also presentation
  by A. Falcone at the Sixth International Fermi Symposium, {\tt http://fermi.gsfc.nasa.gov/science/mtgs/symposia/2015/}.}
While in some cases the {\it Swift} observations have been superseded by observations with more sensitive instruments (e.g. 3FGL J0212.1+5320,
see previous paragraphs), we nevertheless leave the {\it Swift}
results in Table~\ref{counterparts} for completeness, and possible comparison.

\appendix
\section{Scripts}
The following scripts reproduce the key results presented in the
previous sections. They can also be obtained at:
\newline
{\tt http://www.physics.hku.hk/$\sim$pablo/pulsarness.html} 
\newline
and/or:
\newline
 {\tt http://scipp.ucsc.edu/$\sim$pablo/pulsarness.html}.
The scripts were run on R version 3.2.3, on a Mac running OS X,
Version 10.9.5. The following R packages were used: fermicatsR (v1.3),
dplyr (v0.4.3), pROC (v1.8), randomForest (v4.6-12), RWeka(v0.4-24),
e1071 (v1.6-7), ISLR (v1.0), leaps (v2.9), gam (v1.12), and mgcv (v1.8-9).

\subsection{Getting and Cleaning Data}\label{getdata}
\lstset{basicstyle=\footnotesize, language=R, frame=single,
   columns=flexible, upquote}
\begin{lstlisting}[frame = single]
# Load fermicatsR package containing Fermi LAT Catalogs
library(fermicatsR)

# Load dplyr package for efficient handling of data frames
library(dplyr)

# Define function to compute SED points (energy flux at the geometric mean of the energy band)
sedflux <- function(photon_flux, alpha, Elo, Ehi) {
        R <- Elo/Ehi
        GeV2erg <- 0.00160217657
        (GeV2erg*Elo)*(alpha-1)*(photon_flux)*(R^(alpha/2 - 1))/(1-(R^(alpha-1)))
}

# Subselect small number of variables from those available in 3FGL
FGL3_01 <- select(FGL3, 
                  Source_Name, RAJ2000, DEJ2000, GLON, GLAT, 
                  Spectral_Index, Energy_Flux100, Variability_Index, 
                  Conf_68_SemiMajor, Conf_68_SemiMinor, Conf_68_PosAng, 
                  Conf_95_SemiMajor, Conf_95_SemiMinor, Conf_95_PosAng, 
                  Signif_Avg, Pivot_Energy, Flux_Density, Unc_Flux_Density, 
                  Flux1000, Unc_Flux1000, Energy_Flux100, Unc_Energy_Flux100, 
                  Signif_Curve, Flux100_300, Flux300_1000, Flux1000_3000, 
                  Flux3000_10000, Flux10000_100000, CLASS1, ASSOC1)

# Add new variables (SED points and hardness ratios) computed using sedflux function above        
FGL3_02 <- mutate(FGL3_01, 
                  SED100_300 = sedflux(Flux100_300, Spectral_Index, 0.1, 0.3), 
                  SED300_1000 = sedflux(Flux300_1000, Spectral_Index, 0.3, 1.0), 
                  SED1000_3000 = sedflux(Flux1000_3000, Spectral_Index, 1.0, 3.0), 
                  SED3000_10000 = sedflux(Flux3000_10000, Spectral_Index, 3.0, 10.0), 
                  SED10000_100000 = sedflux(Flux10000_100000, Spectral_Index, 10.0, 100.0), 
                  hr12 = (SED300_1000-SED100_300)/(SED300_1000+SED100_300), 
                  hr23 = (SED1000_3000-SED300_1000)/(SED1000_3000+SED300_1000), 
                  hr34 = (SED3000_10000-SED1000_3000)/(SED3000_10000+SED1000_3000), 
                  hr45 = (SED10000_100000-SED3000_10000)/(SED10000_100000+SED3000_10000), 
                  CLASS1 = gsub(" ", "", CLASS1))

# Drop highly correlated and some unused variables
FGL3_03 <- select(FGL3_02, 
                  -Unc_Flux1000, -Energy_Flux100, -Conf_95_SemiMajor, -Conf_95_SemiMinor, 
                  -Flux100_300, -Conf_68_SemiMajor, -Conf_68_SemiMinor, -Conf_68_PosAng, 
                  -Flux1000, -SED100_300, -SED300_1000, -SED1000_3000, -SED3000_10000, 
                  -SED10000_100000, -Flux300_1000, -Flux1000_3000, -Flux3000_10000)

# Take log of variables with highly skewed distributions (removing zeroes to avoid -Inf)
FGL3_04 <-  filter(FGL3_03, FGL3_03$Signif_Curve != 0) %>%
                mutate(Variability_Index = log(Variability_Index), 
                Pivot_Energy = log(Pivot_Energy), 
                Flux_Density = log(Flux_Density), 
                Unc_Flux_Density = log(Unc_Flux_Density), 
                Unc_Energy_Flux100 = log(Unc_Energy_Flux100), 
                Flux10000_100000 = log(Flux10000_100000), 
                Signif_Curve = log(Signif_Curve))

# Drop a few more unused variables
FGL3_05 <- select(FGL3_04, -Pivot_Energy, -Unc_Flux_Density, -Flux10000_100000)

# Create a tidy data set including "pulsarness" and "agnness" factors
FGL3_tidy <- select(FGL3_05, -RAJ2000, -DEJ2000, -GLON, -Signif_Avg) %>%
                mutate(agnness = factor(CLASS1 == "BCU" | CLASS1 == "bcu" 
                                     |CLASS1 == "BLL" | CLASS1 == "bll"
                                     |CLASS1 == "FSRQ"| CLASS1 == "fsrq"
                                     |CLASS1 == "rdg" | CLASS1 == "RDG"
                                     |CLASS1 == "nlsy1" | CLASS1 == "NLSY1"
                                     |CLASS1 == "agn" | CLASS1 == "ssrq"
                                     |CLASS1 == "sey", 
                                        labels = c("Non-AGN", "AGN")),
                        pulsarness = factor(CLASS1 == "PSR" | CLASS1 == "psr",
                                        labels = c("Non-Pulsar", "Pulsar")))

# Create the FGL3_results data frame that will contain final results
FGL3_results <- select(FGL3_05, Source_Name,
                       Signif = Signif_Avg,
                       Flux = Flux_Density,
                       RA = RAJ2000,
                       DEC = DEJ2000,
                       GLON, GLAT, ASSOC1, CLASS1) %>%
        mutate(Source_Name = substr(Source_Name, 6, 18),
               RA = format(RA, digits = 2),
               DEC = format(DEC, digits = 2),
               GLON = format(GLON, digits = 2),
               GLAT = format(GLAT, digits = 2),
               Signif = round(Signif, digits = 3)) %>%
        mutate(Flux = format(exp(Flux), scientific = TRUE, digits = 3))

# Select data for PSR (pulsarness == "Pulsar") vs AGN (agnness == "AGN") classification
FGL3_AGNPSR <- filter(FGL3_tidy, pulsarness == "Pulsar" | agnness == "AGN")

# Drop sources with missing values and some unused variables
FGL3_AGNPSR <- na.omit(FGL3_AGNPSR)
FGL3_AGNPSR <- select(FGL3_AGNPSR, -CLASS1, -ASSOC1, -Source_Name,
-GLAT, -Conf_95_PosAng, -agnness)

# Set Random seed
set.seed(1)

# Separate into training (70%) and testing (30%) sets
train <- sample(nrow(FGL3_AGNPSR), round(nrow(FGL3_AGNPSR)*0.7))
FGL3_test <- FGL3_AGNPSR[-train, ]
FGL3_train <- FGL3_AGNPSR[train, ]

# Split training data into k (10) blocks for k-fold cross-validation
k <- 10
Index <- sample(nrow(FGL3_train))
Block_index <- matrix(data = Index[1:(floor(nrow(FGL3_train)/k)*k)], nrow = k, byrow = TRUE)
FGL3_train_CV <- FGL3_train[Index[1:(floor(nrow(FGL3_train)/k)*k)], ]

# Initialise predictions vector
predictions_FGL3_train_CV <- rep(0, nrow(FGL3_train_CV))

# Prepare pulsar data sets for YNG vs MSP classification
pulsars_long <- mutate(pulsars, PSR_coords = substr(PSR, 6, 12))
FGL3_pulsars <- FGL3_tidy %>%
        filter(pulsarness == "Pulsar") %>%
        mutate(ASSOC1 = as.character(ASSOC1))

a <- strsplit(FGL3_pulsars$ASSOC1, "PSR J")
b <- character(nrow(FGL3_pulsars))

for (i in 1:nrow(FGL3_pulsars)) {
        b[i] <- gsub(" ", "", a[[i]][2])       
}

FGL3_pulsars$ASSOC1_code <- b
FGL3_pulsars <- mutate(FGL3_pulsars, ASSOC1_code = substr(ASSOC1_code, 1, 7))
bothFGL3andpulsars <- inner_join(pulsars_long, FGL3_pulsars, 
                                 by = c("PSR_coords" = "ASSOC1_code"))
FGL3_Pulsars <- mutate(bothFGL3andpulsars, 
                       pulsarness = factor(P_ms >= 10, labels = c("MSP", "YNG"))) %>%
        select(-P_ms, -Codes, -RAJ_deg, -DECJ_deg, -Refs, -CLASS1, -ASSOC1, 
               -Source_Name, -PSR, -PSR_coords, -Edot, -agnness)

# Separate pulsar data set into 30% (testing test) and 70% (training test)
set.seed (1)

train <- sample(nrow(FGL3_Pulsars), round(nrow(FGL3_Pulsars)*0.7))
FGL3_Pulsars_test <- FGL3_Pulsars[-train, ]
FGL3_Pulsars_train <- FGL3_Pulsars[train, ]

# Clean up and Set aside required data sets                                                                                                                      
Environ <- ls()
Environ <- Environ[Environ != "FGL3_tidy"
                   & Environ != "FGL3_results"  
                   & Environ != "FGL3_test" 
                   & Environ != "FGL3_train" 
                   & Environ != "predictions_FGL3_train_CV" 
                   & Environ != "Block_index" 
                   & Environ != "FGL3_train_CV"
                   & Environ != "FGL3_Pulsars_train"
                   & Environ != "FGL3_Pulsars_test"]
rm(list = Environ)

# Save current workspace for subsequent steps
save.image()

\end{lstlisting}

\subsection{Function to get best threshold, plot ROC curves and print tables}\label{roc}
 \lstset{basicstyle=\footnotesize, language=R}
\begin{lstlisting}

# Load workspace from previous step
load(".RData")

# Load pROC package
library(pROC)

# Function to get the best threshold
ROC_threshold <- function(truth, prediction) {
        ROC <- roc(truth, prediction)
        ROC_table <- cbind(ROC$thresholds, ROC$specificities, ROC$sensitivities)
        ROC_table[which.max(ROC_table[, 2] + ROC_table[, 3]), ]        
}

# Function to plot ROC curves
ROC_plots <- function(truth_train, prediction_train, truth_test, prediction_test) {
        par(mfrow = c(1, 2))
                ROC_train <- roc(truth_train, prediction_train, 
                        plot = TRUE, print.auc = TRUE, main = "ROC Train", print.thres = "best")
                ROC_test <- roc(truth_test, prediction_test, 
                        plot = TRUE, print.auc = TRUE, main = "ROC Test", print.thres = "best")
}

# Function to generate contingency tables
ROC_tables <- function(truth_train, prediction_train, truth_test, prediction_test, 
                       cat1 = "Pulsar", cat2 = "AGN") {
        ROC <- roc(truth_train, prediction_train)
        ROC_table <- cbind(ROC$thresholds, ROC$specificities, ROC$sensitivities)
        threshold <- ROC_table[which.max(ROC_table[, 2] + ROC_table[, 3]), ] 

        # Training data
        nrow_train <- length(prediction_train)
        Predict_class_train <- rep("NA", nrow_train)
        Predict_class_train <- ifelse(prediction_train > threshold[1], cat1, cat2) 
        real_category <- truth_train
        print(table(Predict_class_train, real_category))
        
        # Testing data
        nrow_test <- length(prediction_test)
        Predict_class_test <- rep("NA", nrow_test)
        Predict_class_test <- ifelse(prediction_test > threshold[1], cat1, cat2) 
        real_category <- truth_test
        print(table(Predict_class_test, real_category))
}

ROC_threshold_plots_tables <- function(truth_train, prediction_train, truth_test, prediction_test, 
                                       threshold = 0, cat1 = "Pulsar", cat2 = "AGN") {
        
        if (threshold == 0) {
                # Compute best threshold if none is provided
                ROC <- roc(truth_train, prediction_train)
                ROC_table <- cbind(ROC$thresholds, ROC$specificities, ROC$sensitivities)
                best_threshold <- ROC_table[which.max(ROC_table[, 2] + ROC_table[, 3]), ] 
        } else {
                # Don't compute threshold if one is provided
                best_threshold <- threshold
        }

        if (nargs() > 2) { 
                # Make ROC plots
                par(mfrow = c(1, 2))
                ROC_train <- roc(truth_train, prediction_train, 
                         plot = TRUE, print.auc = TRUE, main = "ROC Train", print.thres = "best")
                ROC_test <- roc(truth_test, prediction_test, 
                        plot = TRUE, print.auc = TRUE, main = "ROC Test", print.thres = "best")
        
                # Generate contingency tables
                # Training data
                nrow_train <- length(prediction_train)
                Predict_class_train <- rep("NA", nrow_train)
                Predict_class_train <- ifelse(prediction_train > best_threshold[1], cat1, cat2) 
                real_category <- truth_train
                print(table(Predict_class_train, real_category))

                # Testing data
                nrow_test <- length(prediction_test)
                Predict_class_test <- rep("NA", nrow_test)
                Predict_class_test <- ifelse(prediction_test > best_threshold[1], cat1, cat2) 
                real_category <- truth_test
                print(table(Predict_class_test, real_category))
        }
        return(best_threshold)
}

# Function to get best threshold, plot ROC curves and print tables
ROCandTable <- function(predictions_FGL3_train, FGL3_train, 
predictions_FGL3_test, FGL3_test, Best_threshold) {

        # Plot ROC curve and Table for train and testing data
        par(mfrow = c(1, 2))
        # Draw ROC for train_data
        ROC_FGL3_train <- roc(FGL3_train$pulsarness, predictions_FGL3_train, 
                              plot = TRUE, print.auc = TRUE, main = "ROC FGL3 Train", 
                              print.thres = "best")
        # Draw ROC for test_data
        ROC_FGL3_test <- roc(FGL3_test$pulsarness, predictions_FGL3_test, 
                              plot = TRUE, print.auc = TRUE, main="ROC FGL3 Test", 
                              print.thres = "best")
       
        # Using The Best Threshold from cross-validation method
        Best_threshold 
        # Table for training data
        nrow_train <- nrow(FGL3_train)
        Predict_class <- rep("NA", nrow_train)
        for (i in 1:nrow_train) {
                if (predictions_FGL3_train[i] <= Best_threshold[1]) {
                        Predict_class[i] <- "AGN"
                } else {
                        Predict_class[i] <- "Pulsar"
                }
        }
        real_category <- FGL3_train$pulsarness
        print(table(Predict_class, real_category))
        # Table for testing data
        nrow_test <- nrow(FGL3_test)
        Predict_class <- rep("NA", nrow_test)
        for (i in 1:nrow_test) {
                if (predictions_FGL3_test[i] <= Best_threshold[1]){
                        Predict_class[i] <- "AGN"
                } else {
                        Predict_class[i] <- "Pulsar"
                }
        }
        real_category <- FGL3_test$pulsarness
        print(table(Predict_class, real_category))
}

# Save current workspace for subsequent steps
save.image()

\end{lstlisting}

\subsection{AGN vs PSR classification using Logistic Regression (LR)
  model (backward stepwise selection)}\label{AGNPSRLR}
 \lstset{basicstyle=\footnotesize, language=R}
\begin{lstlisting}
# AGN vs PSR classification using Logistic Regression (LR) model (backward stepwise selection)
# Load workspace from previous step
load(".RData")

# Load pROC package
library(pROC)

null <- glm(pulsarness ~1, family = binomial, data = FGL3_train)
glm.step.backward.AIC <- step(glm(pulsarness ~., family = binomial, data = FGL3_train),
                              scope = list(lower = null, upper = glm(pulsarness ~., 
                              family = binomial, data = FGL3_train)),
                              direction = "backward")
print(summary(glm.step.backward.AIC))

predictions_FGL3_train <- predict(glm.step.backward.AIC, FGL3_train, type = "response")
predictions_FGL3_test <- predict(glm.step.backward.AIC, FGL3_test, type = "response")

# Get Training Set threshold, generate ROC plots, and print contingency tables
Best_threshold_train <- ROC_threshold_plots_tables(FGL3_train$pulsarness, 
                                                   predictions_FGL3_train, 
                                                   FGL3_test$pulsarness, 
                                                   predictions_FGL3_test)

# Add LR Prediction Probabilities to FGL3_results data frame:
FGL3_results$LR_P <- round(predict(glm.step.backward.AIC, FGL3_tidy,
                                   type = "response"), digits = 3)

# Add LR Prediction category to FGL3_results data frame:
FGL3_results$LR_Pred <- ifelse(FGL3_results$LR_P > Best_threshold_train[1],
                               "PSR", "AGN")

# Clean up and Set aside required data sets
Environ <- ls()
Environ <- Environ[Environ != "FGL3_tidy"
                   & Environ != "FGL3_results"
                   & Environ != "FGL3_test" 
                   & Environ != "FGL3_train" 
                   & Environ != "predictions_FGL3_train_CV" 
                   & Environ != "Block_index" 
                   & Environ != "FGL3_train_CV"
                   & Environ != "ROC_threshold_plots_tables"
                   & Environ != "FGL3_Pulsars_train"
                   & Environ != "FGL3_Pulsars_test"]
rm(list = Environ)

# Save current workspace for subsequent steps
save.image()
\end{lstlisting}

\subsection{AGN vs PSR classification using Random Forests (RF)}\label{AGNPSRRF}
 \lstset{basicstyle=\footnotesize, language=R}
\begin{lstlisting}
# AGN vs PSR classification using Random Forests (RF)
# Load workspace from previous step
load(".RData")

# Load randomForest package
library(randomForest)

set.seed(1)

# First use 10-fold cross validation method to fit models and get the forecast of each block of data
k <- 10
for (i in 1:k) {
  Model <- randomForest(pulsarness~., data = FGL3_train[-Block_index[i, ], ], importance = TRUE)
  predictions_FGL3_train_CV[((i-1)*(nrow(FGL3_train_CV)/k) + 1):((i)*(nrow(FGL3_train_CV)/k))] <-
    predict(Model, newdata = FGL3_train[Block_index[i, ], ], type = "Prob")[, 2]
  print(i)
}

# Get best threshold 
Best_threshold_train_CV <- ROC_threshold_plots_tables(FGL3_train_CV$pulsarness, 
predictions_FGL3_train_CV)

# Now Modeling using all the data
rf.full.FGL3 <- randomForest(pulsarness~., data = FGL3_train, importance = TRUE)
# The importance of each variable
importance(rf.full.FGL3)
varImpPlot(rf.full.FGL3)
# Prediction for train and test
predictions_FGL3_train <- predict(rf.full.FGL3, newdata = FGL3_train, type = "Prob")[, 2]
predictions_FGL3_test <- predict(rf.full.FGL3, newdata = FGL3_test, type = "Prob")[,2 ]

# Generate ROC plots, and print contingency tables
ROC_threshold_plots_tables(FGL3_train$pulsarness, predictions_FGL3_train, 
FGL3_test$pulsarness, predictions_FGL3_test, threshold = Best_threshold_train_CV[1])
cat("Best threshold from cross-validation:", Best_threshold_train_CV)

# Add RF Prediction probabilities to FGL3_results
FGL3_results$RF_P <- predict(rf.full.FGL3, newdata = FGL3_tidy, type = "Prob")[,2 ]
#FGL3_results <- mutate(FGL3_results, RF_P)

# Add RF Prediction category to FGL3_results
FGL3_results$RF_Pred <- ifelse(FGL3_results$RF_P > Best_threshold_train_CV[1], 
                                     c("PSR"), c("AGN")) 

# Plot for the paper
par(mfrow = c(1,1))
varImpPlot(rf.full.FGL3, pch = 19, type = 1, main = "")

# Clean up and Set aside required data sets
Environ <- ls()
Environ <- Environ[Environ != "FGL3_tidy"
                   & Environ != "FGL3_results"
                   & Environ != "FGL3_test" 
                   & Environ != "FGL3_train" 
                   & Environ != "predictions_FGL3_train_CV" 
                   & Environ != "Block_index" 
                   & Environ != "FGL3_train_CV"
                   & Environ != "ROC_threshold_plots_tables"
                   & Environ != "FGL3_Pulsars_train"
                   & Environ != "FGL3_Pulsars_test"
                   & Environ != "rf.full.FGL3"]
rm(list = Environ)

# Save current workspace for subsequent steps
save.image()
\end{lstlisting}

\subsection{Outlyingness}\label{outlyingness}
\lstset{basicstyle=\footnotesize, language=R}
\begin{lstlisting}
# Computation of AGN and PSR "outlyingness" using Random Forests (RF)
# Load workspace from previous step
load(".RData")

# Load randomForest package
library(randomForest)

# Set random seed
set.seed(1)

# Compute RF predictions on entire FGL3 data
results <- predict(rf.full.FGL3, newdata = FGL3_tidy, proximity = TRUE, type = "Prob")

# Use proximity matrix to compute PSR and AGN "outlyingness"
nobs <- dim(FGL3_tidy)[1]
PSR_Out <- numeric(nobs)
AGN_Out <- numeric(nobs)

for (i in 1:nobs ) {
        PSR_proximities <- results$proximity[FGL3_tidy$pulsarness == "Pulsar", i] 
        PSR_Out[i] <- 1./sum(PSR_proximities^2)
        AGN_proximities <- results$proximity[FGL3_tidy$agnness == "AGN", i] 
        AGN_Out[i] <- 1./sum(AGN_proximities^2)
}

# Normalize outlyingness      
PSR_Out_norm <- (PSR_Out - median(PSR_Out[FGL3_tidy$pulsarness == "Pulsar"])) / 
        mad(PSR_Out[FGL3_tidy$pulsarness=="Pulsar"])
AGN_Out_norm <- (AGN_Out - median(AGN_Out[FGL3_tidy$agnness == "AGN"])) / 
        mad(AGN_Out[FGL3_tidy$agnness=="AGN"])

# Add outlyingness to FGL3_results data frame
FGL3_results$PSR_Out <- round(PSR_Out_norm, digits = 3)
FGL3_results$AGN_Out <- round(AGN_Out_norm, digits = 3)

# Clean up and Set aside required data sets
Environ <- ls()
Environ <- Environ[Environ != "FGL3_tidy"
                   & Environ != "FGL3_results"
                   & Environ != "FGL3_test" 
                   & Environ != "FGL3_train" 
                   & Environ != "predictions_FGL3_train_CV" 
                   & Environ != "Block_index" 
                   & Environ != "FGL3_train_CV"
                   & Environ != "ROC_threshold_plots_tables"
                   & Environ != "FGL3_Pulsars_train"
                   & Environ != "FGL3_Pulsars_test"]
rm(list = Environ)

# Save current workspace for subsequent steps
save.image()  
\end{lstlisting}

\subsection{Pulsar (YNG vs MSP) classification using Boosted LR}\label{YNGMSPLR}
 \lstset{basicstyle=\footnotesize, language=R}
\begin{lstlisting}
# Pulsar (YNG vs MSP) classification using Boosted LR
# Load workspace from previous step
load(".RData")

# Load RWeka and pROC packages
library(RWeka)
library(pROC)

# WOW(LogitBoost)
Model <- LogitBoost(pulsarness ~., data = FGL3_Pulsars_train)
summary(Model)

# Model predictions
predictions_FGL3_Pulsars_train <- predict(Model, 
                                          newdata = FGL3_Pulsars_train, type = "probability")[, 2]
predictions_FGL3_Pulsars_test <- predict(Model, 
                                         newdata = FGL3_Pulsars_test, type = "probability")[, 2]

# Generate ROC plots and print contingency tables
Best_threshold_train <- ROC_threshold_plots_tables(FGL3_Pulsars_train$pulsarness, 
                                                   predictions_FGL3_Pulsars_train, 
                                                   FGL3_Pulsars_test$pulsarness, 
                                                   predictions_FGL3_Pulsars_test, 
                                                   cat1 = "YNG", cat2 = "MSP")

# Add Boosted Logistic Regression Prediction probabilities to FGL3_results data frame
FGL3_results$Blogistic_YNG_P <- round(predict(Model, 
                                        newdata = FGL3_tidy, type = "probability")[, 2],
                                      digits = 3)

# Add Boosted Logistic Regression Prediction category to FGL3_results data frame
FGL3_results$Blogistic_YNG_Pred <- ifelse(FGL3_results$Blogistic_YNG_P > Best_threshold_train[1], 
                                   c("YNG"), c("MSP")) 

# Clean up and Set aside required data sets
Environ <- ls()
Environ <- Environ[Environ != "FGL3_tidy"
                   & Environ != "FGL3_results"
                   & Environ != "FGL3_test" 
                   & Environ != "FGL3_train" 
                   & Environ != "predictions_FGL3_train_CV" 
                   & Environ != "Block_index" 
                   & Environ != "FGL3_train_CV"
                   & Environ != "ROC_threshold_plots_tables"
                   & Environ != "FGL3_Pulsars_train"
                   & Environ != "FGL3_Pulsars_test"]
rm(list = Environ)

# Save current workspace for subsequent steps
save.image()  
\end{lstlisting}

\subsection{Pulsar (YNG vs MSP) classification using Random Forests (RF)}\label{YNGMSPRF}
 \lstset{basicstyle=\footnotesize, language=R}
\begin{lstlisting}
# Pulsar (YNG vs MSP) classification using Random Forests (RF)
# Load workspace from previous step
load(".RData")

# Load randomForest and pROC packages
library(randomForest)
library(pROC)

# Set random seed
set.seed(1)

# Full model
rf.P <- randomForest(pulsarness ~., data = FGL3_Pulsars_train, importance = TRUE)

# The importance of each variable
importance(rf.P)
varImpPlot(rf.P)

# Model predictions
predictions_FGL3_Pulsars_train <- predict(rf.P, 
                                          newdata = FGL3_Pulsars_train, type = "Prob")[,2 ]
predictions_FGL3_Pulsars_test <- predict(rf.P, 
                                         newdata = FGL3_Pulsars_test, type = "Prob")[,2 ]

# Generate ROC plots and print contingency tables
Best_threshold_train <- ROC_threshold_plots_tables(FGL3_Pulsars_train$pulsarness, 
                                                   predictions_FGL3_Pulsars_train, 
                                                   FGL3_Pulsars_test$pulsarness, 
                                                   predictions_FGL3_Pulsars_test, 
                                                   cat1 = "YNG", cat2 = "MSP")

# Add RF Prediction probabilities to FGL3_results data frame
FGL3_results$RF_YNG_P <- round(predict(rf.P, 
                                       newdata = FGL3_tidy, type = "Prob")[,2 ],
                               digits = 3)

# Add RF Prediction category to FGL3_results data frame
FGL3_results$RF_YNG_Pred <- ifelse(FGL3_results$RF_YNG_P > Best_threshold_train[1], 
                               c("YNG"), c("MSP")) 

# Plot Importance
par(mfrow = c(1,1))
varImpPlot(rf.P, pch = 19, type = 1, main = "")

# Clean up and Set aside required data sets
Environ <- ls()
Environ <- Environ[Environ != "FGL3_tidy"
                   & Environ != "FGL3_results"
                   & Environ != "FGL3_test" 
                   & Environ != "FGL3_train" 
                   & Environ != "predictions_FGL3_train_CV" 
                   & Environ != "Block_index" 
                   & Environ != "FGL3_train_CV"
                   & Environ != "ROC_threshold_plots_tables"
                   & Environ != "FGL3_Pulsars_train"
                   & Environ != "FGL3_Pulsars_test"]
rm(list = Environ)

# Save current workspace for subsequent steps
save.image()  

# Write results to a file
save(FGL3_results, file = "FGL3_results.rdata", compress = "xz")

\end{lstlisting}

\begin{acknowledgments}

We thank the anonymous referee for useful feedback on our original manuscript, helping us to significantly improve the final version.
This work was supported by the National Aeronautics and Space Administration (NASA) through Fermi Guest Investigator grants
NNX10AP18G, NNX12AP41G, and NNG12PP66P. Additional support was provided through NASA Grant and Cooperative Agreement NNX13AB52G and 
Chandra Awards G03-14070X and G03-14079X, issued by the Chandra X-ray
Observatory Center, which is operated by the Smithsonian Astrophysical Observatory for and on behalf of NASA under contract NAS8-03060.

\end{acknowledgments}


\bibliographystyle{apj}
\bibliography{pulsarness}


\begin{deluxetable}{lrrr}
\tablewidth{0pt}
\tablecaption{Predictor Parameters used for the AGN vs PSR
  models\tablenotemark{a}. \label{variables}}
\tablehead{
\colhead{Parameter} & \colhead{Min.} & \colhead{Median} & \colhead{Max.} \\
}
\startdata
Spectral\_Index  & 0.5 & 2.2 & 3.1 \\
Variability\_Index\tablenotemark{b} & 3.0 & 4.0 & 11.0\\
Flux\_Density\tablenotemark{c}  & -35.4 & -28.2 & -19.9 \\
Unc\_Energy\_Flux100\tablenotemark{d} & -28.5 & -27.6 & -24.8 \\  
Signif\_Curve\tablenotemark{b}  & -5.8 & 0.4 & 4.4 \\
$hr_{12}$ & -1 & -0.1 & 1 \\
$hr_{23}$ & -1 & -0.1 & 1 \\
$hr_{34}$ & -1 & -0.2 & 1 \\ 
$hr_{45}$ & -1 & -0.3 & 1 \\
\enddata
\tablenotetext{a}{For the YNG vs MSP models we also used the Galactic
  latitude (GLAT) of the source, as a
  predictor parameter.}
\tablenotetext{b}{Number represents the log of the original value
  contained in the catalog.}
\tablenotetext{c}{In photon cm$^{-2}$ MeV$^{-1}$ s$^{-1}$ (log of the original value
  contained in the catalog).}
\tablenotetext{d}{In erg cm$^{-2}$ s$^{-1}$ (log of the original value
  contained in the catalog).}
\end{deluxetable}

\begin{deluxetable}{llllll}
\tablewidth{0pt}
\tablecaption{Performance of the algorithms in classifying AGN vs PSR, based on a testing sample
  of 521 AGN and 50 PSR (in bold are the two best methods, Random
  Forest (RF) and Logistic Regression (LR), as
  described in Sections \ref{RFsection} and \ref{LRsection})).\label{AGNvPSR}}
\tablehead{
\colhead{Model} & \colhead{AGN} & \colhead{PSR} & \colhead{PSR}  &
\colhead{Overall} & F1 \\
  & \colhead{Test Errors} & \colhead{Test Errors}
  &\colhead{Sensitivity} & \colhead{Accuracy} & Score \\
  & \colhead{(out of 521)} & \colhead{(out of 50)} &\colhead{} &
  \colhead{} & \colhead{} 
}
\startdata
LR (forward stepwise)  & 38 & 1 & 98\% & 93.2\% & 71.5\% \\
{\bf LR (backward stepwise)}  & {\bf 29} & {\bf 1} & {\bf 98\%} & {\bf
  94.7\%}  & 76.6\% \\
Decision Tree (C4.5)  & 15 & 6 & 88\% & 96.3\% & 80.7\% \\
Two-stage\tablenotemark{a}  & 17 & 6 & 88\% & 96.0\% & 79.3\% \\
GAM\tablenotemark{b}  & 25 & 4 & 92\% & 94.9\% & 76.0\% \\
SVM (CV)\tablenotemark{c}  & 29 & 1 & 98\% & 94.8\% & 76.6\% \\
LR (CV)\tablenotemark{d}  & 31 & 1 &  98\% & 94.4\% & 75.4\% \\
Boosted LR (CV)  & 24 & 5 & 90\% & 94.9\% & 75.6\% \\
Logistic Trees (CV)  & 36 & 2 & 96\% & 93.4\% & 71.6\% \\
Decision Tree (C4.5 with CV)  & 15 & 6 & 88\% & 96.3\% & 80.7\% \\
{\bf RF (CV)}\tablenotemark{j}  & {\bf 17} & {\bf 2} & {\bf  96\%} &
{\bf 96.7\%} & 83.5\% \\

\enddata
\tablenotetext{a}{Decision Tree (C4.5) + logistic regression.}
\tablenotetext{b}{General Additive Model}
\tablenotetext{c}{Support Vector Machine. Here, and in all other
  cases, CV refers to the use of 10-fold cross-validation.}
\tablenotetext{d}{With either forward or backward stepwise elimination.}
\end{deluxetable}

\begin{deluxetable}{llll}
\tablewidth{0pt}
\tablecaption{Performance of the algorithms in classifying Young (YNG)
  vs Millisecond pulsars (MSP), based on a testing sample
  of 25 YNG and 18 MSP (in bold are the two best methods, RF and Boosted LR, described in
  Sections \ref{RFsection} and \ref{BLRsection}).\label{YNGvMSP}}
\tablehead{
\colhead{Model} & \colhead{YNG} & \colhead{MSP} & \colhead{Overall} \\
  & \colhead{Test Errors} & \colhead{Test Errors} & \colhead{Accuracy}\\
  & \colhead{(out of 25)} & \colhead{(out of 18)} & \colhead{}
}
\startdata

LR (forward stepwise)  & 7 & 2 & 79.1\% \\
LR (backward stepwise)  & 7 & 2 & 79.1\%  \\
{\bf Boosted LR}  & {\bf 2} & {\bf 2} & {\bf 90.7\%} \\
Logistic Trees & 7 & 2 & 79.1\% \\
GAM  & 2 & 4 & 86.1\% \\
SVM  & 4 & 3 & 83.7\% \\
Decision Tree (C4.5)  & 7 & 0 & 83.7\% \\
{\bf RF} & {\bf 2} & {\bf 2} & {\bf 90.7\%} \\
\enddata
\end{deluxetable}

\begin{deluxetable}{lrrrr}
\tablewidth{0pt}
\tablecaption{Predictor parameters for the logistic regression model
  (backwards stepwise)\label{logit}}
\tablehead{
\colhead{Parameter} & \colhead{Estimate} & \colhead{Std. Error} & \colhead{z
value} & \colhead{Pr($> |z|$)} \\
}
\startdata

(Intercept) & 151.7880  & 23.5877 &  6.435 & 1.23e-10 \\
Variability\_Index~\tablenotemark{a} &  -5.5473 & 0.9253 & -5.995 & 2.03e-09 \\
Unc\_Energy\_Flux100~\tablenotemark{a} & 3.7512 &  0.6978 & 5.376 & 7.64e-08 \\
hr45 & -4.3291 &  0.9729  & -4.450 & 8.59e-06 \\
hr34 & -4.2981 & 1.3213  & -3.253  & 0.00114 \\
Spectral\_Index & -5.8615 & 1.8620  & -3.148 & 0.00164 \\ 
Flux\_Density~\tablenotemark{a} & 0.7541 & 0.2726 & 2.766 & 0.00567 \\ 
Signif\_Curve~\tablenotemark{a} & 1.2275 & 0.5569 & 2.204 & 0.02750 \\ 
hr23  & 1.7846 & 1.1325 & 1.576 & 0.11506 \\

\enddata
\tablenotetext{a}{Number is the log of the original value contained in the catalog.}
\end{deluxetable}

\rotate
\begin{deluxetable}{lrlllllllrlrlrrrlrl}
\tabletypesize{\footnotesize}
\setlength{\tabcolsep}{0.02in}
\tablewidth{0pt}
\tablecaption{Results of our best models, as applied to the 3FGL
catalog (all 3021 sources for which predictor parameters are
available, for which our models could therefore be applied). The
table is published in its entirety in the electronic edition of the
article. A portion is shown here for guidance regarding its form and content. The columns are:
3FGL Name (Source\_Name), 4-year significance (Signif), Flux Density, in photon/cm2/MeV/s (Flux), RA, DEC,
Galactic Longitude and Latitude (GLON, GLAT), Name of identified or likely associated source (ASSOC1),
class designation from 3FGL catalog (CLASS1), LR Probability and predicted class (LR\_P, Pred), RF Probability
and predicted class (RF\_P, Pred), PSR Outlyingness (PSR\_Out), AGN Outlyingness (AGN\_Out), Boosted LR Probability
(of being a YNG pulsar) and predicted pulsar class (BLR\_P, BLR\_Pred), and RF Probability (of being a YNG pulsar)
and predicted pulsar class (RF\_P and RF\_Pred). \label{allresults}}
\tablehead{
 \colhead{Source\_Name} & \colhead{Signif} & \colhead{Flux} &
       \colhead{RA} & \colhead{DEC} & \colhead{GLON} & \colhead{GLAT} &
               \colhead{ASSOC1} &
\colhead{CLASS1} & \colhead{LR} & \colhead{LR} & \colhead{RF}
        & \colhead{RF} &
\colhead{PSR} & \colhead{AGN} & \colhead{BLR} &
      \colhead{BLR} &
\colhead{RF} & \colhead{RF} \\
 \colhead{3FGL} & & & & & & & & & \colhead{P} & \colhead{Pred} & \colhead{P}
        & \colhead{Pred} &
\colhead{Out} & \colhead{Out} & \colhead{P} &
      \colhead{Pred} &
\colhead{P} & \colhead{Pred}
}
\startdata
J0000.1+6545 & 6.81 & 1.01e-12 &   0.038 &  65.7517 & 117.69 &   3.4030 &                            &  & 0.38 & PSR & 0.19 & PSR & 203.05 & 10.79 & 0.22 & MSP & 0.52 & YNG \\ 
J0000.2-3738 & 5.09 & 1.94e-14 &   0.061 & -37.6484 & 345.41 & -74.9468 &                            &  & 0.00 & AGN & 0.00 & AGN & 109.49 & -0.63 & 0.00 &  & 0.13 &  \\ 
J0001.0+6314 & 6.16 & 8.62e-12 &   0.254 &  63.2440 & 117.29 &   0.9257 &                            & spp & 0.00 & AGN & 0.05 & AGN & 187.32 & 4.20 & 0.16 &  & 0.41 &  \\ 
J0001.2-0748 & 11.25 & 4.85e-13 &   0.321 &  -7.8159 &  89.02 & -67.3242 & PMN J0001-0746             & bll & 0.02 & AGN & 0.01 & AGN & 81.14 & 2.64 & 0.00 &  & 0.04 &  \\ 
J0001.4+2120 & 11.35 & 2.52e-11 &   0.361 &  21.3379 & 107.67 & -40.0472 & TXS 2358+209               & fsrq & 0.00 & AGN & 0.00 & AGN & 203.17 & 1.55 & 0.01 &  & 0.32 &  \\ 
J0001.6+3535 & 4.20 & 2.87e-13 &   0.404 &  35.5905 & 111.66 & -26.1885 &                            &  & 0.00 & AGN & 0.04 & AGN & 194.12 & 0.33 & 0.00 &  & 0.17 &  \\ 
J0002.0-6722 & 5.89 & 4.83e-14 &   0.524 & -67.3703 & 310.14 & -49.0618 &                            &  & 0.00 & AGN & 0.00 & AGN & 100.44 & -0.72 & 0.00 &  & 0.13 &  \\ 
J0002.2-4152 & 5.16 & 7.32e-14 &   0.562 & -41.8828 & 334.07 & -72.1427 & 1RXS J000135.5-415519      & bcu & 0.00 & AGN & 0.00 & AGN & 120.90 & -0.78 & 0.00 &  & 0.16 &  \\ 
J0002.6+6218 & 17.97 & 4.30e-12 &   0.674 &  62.3006 & 117.30 &  -0.0371 &                            &  & 0.99 & PSR & 0.81 & PSR & 1.98 & 751.35 & 0.82 & YNG & 0.54 & YNG \\ 
J0003.2-5246 & 5.67 & 2.00e-14 &   0.815 & -52.7771 & 318.98 & -62.8247 & RBS 0006                   & bcu & 0.00 & AGN & 0.00 & AGN & 148.05 & -0.61 & 0.00 &  & 0.13 &  \\ 
J0003.4+3100 & 6.29 & 2.03e-12 &   0.858 &  31.0085 & 110.96 & -30.7451 &                            &  & 0.15 & PSR & 0.08 & AGN & 181.42 & 15.50 & 0.01 & MSP & 0.17 & MSP \\ 
J0003.5+5721 & 5.38 & 1.09e-13 &   0.890 &  57.3597 & 116.49 &  -4.9116 &                            &  & 0.04 & AGN & 0.04 & AGN & 206.61 & 1.63 & 0.01 &  & 0.24 &  \\ 
J0003.8-1151 & 4.18 & 4.42e-14 &   0.959 & -11.8627 &  84.43 & -71.0842 & PMN J0004-1148             & bcu & 0.00 & AGN & 0.02 & AGN & 109.97 & 0.04 & 0.00 &  & 0.12 &  \\ 
J0004.2+6757 & 6.01 & 6.01e-13 &   1.055 &  67.9593 & 118.51 &   5.4940 &                            &  & 0.67 & PSR & 0.54 & PSR & 76.78 & 218.93 & 0.49 & MSP & 0.39 & MSP \\ 
\enddata
\end{deluxetable}

\newpage
\begin{longtable}{rlrllrrl}
  \hline
 & Source\_Name & Signif & RA & DEC & LR\_P & RF\_P & BLR/RF Pred \\ 
  \hline
1 &  3FGL~J1744.1-7619 & 32.85 & 266.045 & -76.3286 & 1.00 & 0.96 & MSP/MSP \\ 
  2 &  {\bf 3FGL~J1653.6-0158} & 31.95 & 253.419 &  -1.9801 & 0.99 & 0.73 & MSP/MSP \\ 
  3 &  3FGL~J1035.7-6720 & 30.48 & 158.926 & -67.3335 & 1.00 & 0.94 & MSP/MSP \\ 
  4 &  3FGL~J1119.9-2204 & 30.16 & 169.984 & -22.0673 & 0.99 & 0.91 & MSP/MSP \\ 
  5 &  3FGL~J2112.5-3044 & 30.14 & 318.145 & -30.7344 & 0.99 & 0.94 & MSP/MSP \\ 
  6 &  3FGL~J1702.8-5656 & 28.81 & 255.720 & -56.9357 & 0.92 & 0.36 & MSP/MSP \\ 
  7 &  3FGL~J1625.1-0021 & 28.53 & 246.279 &  -0.3586 & 1.00 & 0.98 & MSP/MSP \\ 
  8 &  {\bf 3FGL~J1906.6+0720} & 26.27 & 286.671 &   7.3339 & 1.00 & 0.85 & YNG/YNG \\ 
  9 &  {\bf 3FGL~J0523.3-2528} & 26.14 &  80.839 & -25.4763 & 0.52 & 0.36 & MSP/MSP \\ 
  10 &  3FGL~J1306.4-6043 & 25.45 & 196.615 & -60.7317 & 1.00 & 0.90 & MSP/MSP \\ 
  11 & {\bf  3FGL~J2039.6-5618} & 25.38 & 309.918 & -56.3121 & 0.99 & 0.82 & MSP/MSP \\ 
  12 &  3FGL~J0212.1+5320 & 25.07 &  33.037 &  53.3360 & 0.98 & 0.92 & MSP/MSP \\ 
  13 &  3FGL~J2017.9+3627 & 25.04 & 304.485 &  36.4591 & 1.00 & 0.85 & YNG/YNG \\ 
  14 &  3FGL~J1405.4-6119 & 24.82 & 211.356 & -61.3168 & 1.00 & 0.77 & YNG/YNG \\ 
  15 &  3FGL~J0340.4+5302 & 22.31 &  55.100 &  53.0484 & 0.99 & 0.28 & YNG/YNG \\ 
  16 &  3FGL~J0933.9-6232 & 22.07 & 143.481 & -62.5338 & 0.99 & 0.96 & MSP/MSP \\ 
  17 &  3FGL~J0634.1+0424 & 21.06 &  98.528 &   4.4062 & 0.98 & 0.69 & YNG/YNG \\ 
  18 &  3FGL~J1745.3-2903c & 20.66 & 266.343 & -29.0630 & 1.00 & 0.87 & YNG/YNG \\ 
  19 &  3FGL~J1622.9-5004 & 20.65 & 245.726 & -50.0753 & 0.99 & 0.75 & YNG/YNG \\ 
  20 & {\bf  3FGL~J1946.4-5403} & 20.29 & 296.614 & -54.0570 & 0.98 & 0.96 & MSP/MSP \\ 
  21 &  3FGL~J1747.0-2828 & 20.26 & 266.775 & -28.4819 & 1.00 & 0.73 & YNG/YNG \\ 
  22 &  3FGL~J1624.2-4041 & 19.27 & 246.059 & -40.6865 & 0.99 & 0.85 & YNG/YNG \\ 
  23 &  3FGL~J1539.2-3324 & 19.23 & 234.823 & -33.4142 & 0.96 & 0.93 & MSP/MSP \\ 
  24 &  3FGL~J0359.5+5413 & 19.17 &  59.881 &  54.2220 & 1.00 & 0.83 & MSP/YNG \\ 
  25 &  3FGL~J0954.8-3948 & 18.89 & 148.712 & -39.8087 & 0.86 & 0.26 & MSP/MSP \\ 
  26 &  3FGL~J1848.4-0141 & 18.67 & 282.118 &  -1.6927 & 1.00 & 0.73 & YNG/YNG \\ 
  27 &  3FGL~J2004.4+3338 & 18.46 & 301.103 &  33.6451 & 0.66 & 0.20 & MSP/YNG \\ 
  28 &  3FGL~J2041.1+4736 & 18.38 & 310.281 &  47.6030 & 0.99 & 0.43 & YNG/YNG \\ 
  29 &  3FGL~J0854.8-4503 & 18.21 & 133.711 & -45.0616 & 0.98 & 0.82 & YNG/YNG \\ 
  30 &  3FGL~J0744.1-2523 & 18.10 & 116.045 & -25.3994 & 0.97 & 0.38 & MSP/YNG \\ 
  31 &  3FGL~J0002.6+6218 & 17.97 &   0.674 &  62.3006 & 0.99 & 0.81 & YNG/YNG \\ 
  32 &  3FGL~J0336.1+7500 & 17.75 &  54.044 &  75.0153 & 0.95 & 0.72 & MSP/MSP \\ 
  33 &  3FGL~J1800.8-2402 & 17.48 & 270.222 & -24.0351 & 0.98 & 0.77 & YNG/YNG \\ 
  34 &  3FGL~J1754.0-2538 & 16.84 & 268.508 & -25.6486 & 0.95 & 0.79 & YNG/YNG \\ 
  35 &  3FGL~J1823.2-1339 & 16.64 & 275.820 & -13.6513 & 1.00 & 0.80 & YNG/YNG \\ 
  36 &  3FGL~J1208.4-6239 & 16.23 & 182.120 & -62.6612 & 0.98 & 0.75 & YNG/YNG \\ 
  37 &  3FGL~J1650.3-4600 & 16.18 & 252.599 & -46.0141 & 0.99 & 0.84 & YNG/YNG \\ 
  38 &  3FGL~J1056.7-5853 & 16.11 & 164.179 & -58.8960 & 1.00 & 0.67 & YNG/YNG \\ 
  39 &  3FGL~J1112.0-6135 & 15.69 & 168.017 & -61.5842 & 0.97 & 0.65 & YNG/YNG \\ 
  40 &  3FGL~J0914.5-4736 & 15.57 & 138.641 & -47.6152 & 0.50 & 0.39 & YNG/YNG \\ 
  41 &  3FGL~J1358.5-6025 & 15.43 & 209.643 & -60.4298 & 1.00 & 0.70 & YNG/YNG \\ 
  42 &  3FGL~J1104.9-6036 & 15.40 & 166.248 & -60.6091 & 1.00 & 0.85 & YNG/YNG \\ 
  43 &  3FGL~J0545.6+6019 & 15.39 &  86.415 &  60.3219 & 0.56 & 0.37 & MSP/MSP \\ 
  44 &  3FGL~J1742.6-3321 & 15.31 & 265.665 & -33.3562 & 1.00 & 0.83 & YNG/YNG \\ 
  45 &  3FGL~J1857.9+0210 & 15.19 & 284.490 &   2.1704 & 1.00 & 0.85 & YNG/YNG \\ 
  46 &  3FGL~J1740.5-2843 & 15.13 & 265.125 & -28.7170 & 1.00 & 0.76 & YNG/YNG \\ 
  47 &  3FGL~J1748.3-2815c & 15.11 & 267.092 & -28.2589 & 0.98 & 0.78 & YNG/YNG \\ 
  48 &  3FGL~J1754.0-2930 & 15.03 & 268.500 & -29.5059 & 0.99 & 0.65 & YNG/YNG \\ 
  49 &  3FGL~J0238.0+5237 & 14.99 &  39.505 &  52.6256 & 0.97 & 0.65 & MSP/MSP \\ 
  50 &  3FGL~J1839.3-0552 & 14.92 & 279.848 &  -5.8816 & 1.00 & 0.88 & YNG/YNG \\ 
  51 &  3FGL~J2117.6+3725 & 14.50 & 319.421 &  37.4256 & 0.98 & 0.30 & MSP/MSP \\ 
  52 &  3FGL~J0419.1+6636 & 14.47 &  64.778 &  66.6049 & 0.26 & 0.33 & MSP/MSP \\ 
  53 &  3FGL~J2212.5+0703 & 14.28 & 333.147 &   7.0598 & 0.69 & 0.40 & MSP/MSP \\ 
  54 &  3FGL~J1447.3-5800 & 14.23 & 221.831 & -58.0049 & 1.00 & 0.61 & YNG/YNG \\ 
  55 &  3FGL~J0426.7+5437 & 14.19 &  66.681 &  54.6168 & 1.00 & 0.71 & YNG/YNG \\ 
  56 &  3FGL~J0858.6-4357 & 14.13 & 134.657 & -43.9582 & 0.94 & 0.56 & YNG/YNG \\ 
  57 &  3FGL~J0953.7-1510 & 14.00 & 148.429 & -15.1745 & 0.97 & 0.88 & MSP/MSP \\ 
  58 &  3FGL~J1901.5-0126 & 13.97 & 285.399 &  -1.4481 & 0.31 & 0.26 & YNG/YNG \\ 
  59 &  3FGL~J1624.1-4700 & 13.90 & 246.042 & -47.0058 & 0.88 & 0.58 & YNG/YNG \\ 
  60 &  3FGL~J0838.8-2829 & 13.74 & 129.704 & -28.4892 & 0.60 & 0.57 & MSP/MSP \\ 
  61 &  3FGL~J1852.8+0158 & 13.71 & 283.209 &   1.9722 & 0.98 & 0.78 & YNG/YNG \\ 
  62 &  3FGL~J1317.6-6315 & 13.53 & 199.403 & -63.2598 & 0.97 & 0.64 & YNG/YNG \\ 
  63 &  3FGL~J1231.6-5113 & 13.45 & 187.902 & -51.2210 & 0.99 & 0.39 & YNG/MSP \\ 
  64 &  3FGL~J1641.5-5319 & 13.22 & 250.378 & -53.3237 & 0.99 & 0.57 & YNG/MSP \\ 
  65 &  3FGL~J1120.6+0713 & 13.18 & 170.172 &   7.2235 & 0.11 & 0.45 & MSP/MSP \\ 
  66 &  3FGL~J1026.2-5730 & 13.14 & 156.560 & -57.5166 & 0.96 & 0.85 & YNG/YNG \\ 
  67 &  3FGL~J1843.7-0322 & 13.09 & 280.928 &  -3.3772 & 1.00 & 0.70 & YNG/YNG \\ 
  68 &  3FGL~J0223.6+6204 & 12.88 &  35.906 &  62.0811 & 1.00 & 0.86 & YNG/YNG \\ 
  69 &  3FGL~J0737.2-3233 & 12.87 & 114.314 & -32.5588 & 0.98 & 0.51 & YNG/MSP \\ 
  70 &  3FGL~J1503.5-5801 & 12.80 & 225.893 & -58.0294 & 0.92 & 0.51 & YNG/YNG \\ 
  71 &  3FGL~J0318.1+0252 & 12.76 &  49.536 &   2.8695 & 0.88 & 0.82 & MSP/MSP \\ 
  72 &  3FGL~J1740.5-2726 & 12.75 & 265.134 & -27.4500 & 1.00 & 0.70 & YNG/YNG \\ 
  73 &  3FGL~J2233.1+6542 & 12.69 & 338.279 &  65.7148 & 0.98 & 0.32 & YNG/YNG \\ 
  74 &  3FGL~J1857.2+0059 & 12.39 & 284.310 &   0.9863 & 0.99 & 0.79 & YNG/YNG \\ 
  75 &  3FGL~J2103.7-1113 & 12.33 & 315.942 & -11.2291 & 0.30 & 0.42 & MSP/MSP \\ 
  76 &  3FGL~J0802.3-5610 & 12.31 & 120.583 & -56.1688 & 0.87 & 0.41 & MSP/MSP \\ 
  77 &  3FGL~J1350.4-6224 & 12.31 & 207.622 & -62.4120 & 0.98 & 0.79 & YNG/YNG \\ 
  78 &  3FGL~J1139.0-6244 & 12.23 & 174.751 & -62.7368 & 0.98 & 0.54 & YNG/YNG \\ 
  79 &  3FGL~J0847.4-4327 & 12.20 & 131.863 & -43.4587 & 0.47 & 0.61 & YNG/YNG \\ 
  80 &  3FGL~J0758.6-1451 & 12.06 & 119.656 & -14.8661 & 0.95 & 0.54 & MSP/MSP \\ 
  81 &  3FGL~J0039.3+6256 & 11.99 &   9.834 &  62.9415 & 0.92 & 0.89 & MSP/MSP \\ 
  82 &  3FGL~J2035.0+3634 & 11.92 & 308.758 &  36.5814 & 0.94 & 0.77 & MSP/MSP \\ 
  83 &  3FGL~J2048.8+4436 & 11.89 & 312.224 &  44.6082 & 0.99 & 0.48 & YNG/YNG \\ 
  84 &  3FGL~J1329.8-6109 & 11.85 & 202.468 & -61.1620 & 0.94 & 0.55 & MSP/MSP \\ 
  85 &  3FGL~J1813.6-1148 & 11.84 & 273.423 & -11.8084 & 1.00 & 0.68 & YNG/YNG \\ 
  86 &  3FGL~J2038.4+4212 & 11.76 & 309.625 &  42.2085 & 0.96 & 0.73 & YNG/YNG \\ 
  87 &  3FGL~J1749.2-2911 & 11.72 & 267.315 & -29.1929 & 1.00 & 0.87 & YNG/YNG \\ 
  88 &  3FGL~J2023.5+4126 & 11.63 & 305.876 &  41.4337 & 0.98 & 0.53 & YNG/YNG \\ 
  89 &  3FGL~J1528.3-5836 & 11.53 & 232.078 & -58.6077 & 0.59 & 0.65 & MSP/MSP \\ 
  90 &  3FGL~J1650.0-4438c & 11.50 & 252.507 & -44.6387 & 0.92 & 0.56 & YNG/YNG \\ 
  91 &  3FGL~J1557.0-4225 & 11.46 & 239.264 & -42.4218 & 0.98 & 0.32 & YNG/YNG \\ 
  92 &  3FGL~J1016.5-6034 & 11.34 & 154.135 & -60.5764 & 0.49 & 0.33 & MSP/MSP \\ 
  93 &  3FGL~J2032.5+3921 & 11.31 & 308.139 &  39.3589 & 0.91 & 0.72 & YNG/YNG \\ 
  94 &  3FGL~J1033.0-5945 & 11.30 & 158.266 & -59.7513 & 0.93 & 0.44 & YNG/YNG \\ 
  95 &  3FGL~J1552.8-5330 & 11.20 & 238.212 & -53.5131 & 1.00 & 0.85 & YNG/YNG \\ 
  96 &  3FGL~J1753.6-4447 & 11.12 & 268.404 & -44.7930 & 0.75 & 0.68 & MSP/MSP \\ 
  97 &  3FGL~J1037.9-5843 & 11.11 & 159.488 & -58.7300 & 0.92 & 0.62 & YNG/YNG \\ 
  98 &  3FGL~J0541.1+3553 & 10.94 &  85.277 &  35.8975 & 0.99 & 0.69 & YNG/YNG \\ 
  99 &  3FGL~J0312.1-0921 & 10.92 &  48.041 &  -9.3593 & 0.74 & 0.52 & MSP/MSP \\ 
  100 &  3FGL~J0855.4-4818 & 10.88 & 133.855 & -48.3163 & 0.99 & 0.68 & YNG/YNG \\ 
  101 &  3FGL~J1919.9+1407 & 10.88 & 289.981 &  14.1179 & 0.52 & 0.68 & YNG/YNG \\ 
  102 &  {\bf 3FGL~J1544.6-1125} & 10.85 & 236.170 & -11.4275 & 0.64 &0.29& MSP/MSP \\ 
  103 &  3FGL~J2042.4+4209 & 10.74 & 310.614 &  42.1532 & 0.62 & 0.52 & YNG/YNG \\ 
  104 &  3FGL~J1737.9-2511 & 10.73 & 264.476 & -25.1858 & 0.97 & 0.54 & YNG/YNG \\ 
  105 &  3FGL~J2333.0-5525 & 10.72 & 353.264 & -55.4265 & 0.10 & 0.18 & MSP/MSP \\ 
  106 &  3FGL~J2039.4+4111 & 10.72 & 309.854 &  41.1982 & 0.94 & 0.70 & YNG/YNG \\ 
  107 &  3FGL~J1616.8-5343 & 10.63 & 244.205 & -53.7243 & 0.88 & 0.43 & MSP/YNG \\ 
  108 &  3FGL~J0857.6-4258 & 10.62 & 134.409 & -42.9824 & 0.76 & 0.50 & YNG/YNG \\ 
  109 &  3FGL~J1518.2-5232 & 10.36 & 229.571 & -52.5349 & 0.36 & 0.35 & MSP/MSP \\ 
  110 &  3FGL~J2034.6+4302 & 10.34 & 308.670 &  43.0390 & 0.99 & 0.74 & YNG/YNG \\ 
  111 &  3FGL~J0641.1+1004 & 10.22 & 100.286 &  10.0833 & 0.61 & 0.24 & YNG/YNG \\ 
  112 &  3FGL~J1652.8-4351 & 10.20 & 253.203 & -43.8566 & 0.96 & 0.67 & YNG/YNG \\ 
  113 &  3FGL~J1627.8+3217 & 10.17 & 246.968 &  32.2988 & 0.39 & 0.38 & MSP/MSP \\ 
  114 &  3FGL~J2133.0-6433 & 10.13 & 323.259 & -64.5535 & 0.12 & 0.53 & MSP/MSP \\ 
  115 &  3FGL~J0915.8-5110 & 10.10 & 138.964 & -51.1799 & 0.68 & 0.22 & YNG/YNG \\ 
  116 &  3FGL~J1900.8+0337 & 10.10 & 285.205 &   3.6230 & 1.00 & 0.69 & YNG/YNG \\ 
  117 &  3FGL~J1844.3-0344 & 10.06 & 281.100 &  -3.7468 & 1.00 & 0.84 & YNG/YNG \\ 
  118 &  3FGL~J0940.6-7609 & 10.06 & 145.164 & -76.1608 & 0.43 & 0.39 & MSP/MSP \\ 
  119 &  3FGL~J1744.7-2252 & 10.05 & 266.181 & -22.8695 & 0.95 & 0.47 & YNG/YNG \\ 
  120 &  3FGL~J1727.7-2637 & 10.04 & 261.946 & -26.6251 & 1.00 & 0.53 & YNG/YNG \\ 
   \hline
\hline
\caption{List of the most significant 3FGL unassociated sources
  predicted by both RF and LR to be pulsars. Column definitions are
  the same as in Table~\ref{allresults}. LR\_P and RF\_P give the
  probabilities of being a pulsar (according to the LR and RF
  algorithms), while the last two columns give the predicted type of
  pulsar (according to the BLR and RF algorithms, respectively). In
  bold we highlight two recently-discovered pulsars and four strong
  millisecond pulsar candidates (see text for details).} 
\label{unassoc}
\end{longtable}

\newpage
\begin{longtable}{llrlrrl}
  \hline
 & Source\_Name & Signif & LAT SNR & LR\_P & RF\_P & BLR/RF Pred \\
  \hline
  1 &  3FGL~J0001.0+6314 & 6.16 &                            & 0.00 & 0.05 & -/-   \\ 
  2 &  3FGL~J0025.7+6404 & 4.41 & Tycho                      & 0.00 & 0.05 &  -/-  \\ 
  3 &  3FGL~J0128.4+6257 & 8.26 &                            & 0.79 & 0.25 & MSP/MSP \\ 
  4 &  3FGL~J0220.1+6202c & 4.63 & G132.7+01.3      & 0.95 & 0.69 & YNG/YNG \\ 
  5 &  3FGL~J0224.0+6235 & 4.57 &                            & 0.33 & 0.26 & YNG/YNG \\ 
  6 &  3FGL~J0454.6-6825 & 6.32 & PWN G279.8-35.8            & 0.20 & 0.19 & MSP/MSP \\ 
  7 &  3FGL~J0500.3+5237 & 8.57 &                            & 0.00 & 0.02 & -/-\\ 
  8 &  3FGL~J0540.3+2756e & 20.93 &  {\bf G180.0--01.7/S147}  & 0.82 & 0.40 & YNG/-  \\ 
  9 &  3FGL~J0610.6+1728 & 10.01 &                            & 0.16 & 0.28 & YNG/YNG \\ 
  10 &  3FGL~J0617.2+2234e & 133.26 &  {\bf G189.1+0.3/IC 443} & 1.00 & 0.74 & YNG/-  \\ 
  11 &  3FGL~J0631.6+0644 & 13.99 & {\bf G205.5+00.5} & 0.83 & 0.70 & YNG/YNG \\ 
  12 &  3FGL~J0640.9+0752 & 5.73 &                            & 0.03 & 0.10 & -/-\\ 
  13 &  3FGL~J0822.6-4250e & 31.71 & {\bf G260.4--03.4/Puppis A} & 0.52 & 0.33 & YNG/-  \\ 
  14 &  3FGL~J0838.1-4615 & 8.66 &                            & 0.90 & 0.62 & YNG/YNG \\ 
  15 &  3FGL~J0839.1-4739 & 4.42 &                            & 0.01 & 0.41 & YNG/YNG \\ 
  16 &  3FGL~J0843.1-4546 & 5.96 &                            & 0.02 & 0.63 & YNG/YNG \\ 
  17 &  3FGL~J0852.7-4631e & 30.28 & {\bf G266.2--01.2/Vela Jr} & 0.06 & 0.28 & YNG/-  \\ 
  18 &  3FGL~J1101.9-6053 & 10.60 &                            & 1.00 & 0.87 & YNG/YNG \\ 
  19 &  3FGL~J1111.9-6038 & 29.19 & {\bf G291.0--00.1}  & 0.98 & 0.78 & YNG/YNG \\ 
  19$^*$ & 3FGL~J1119.1-6127 & 26.97 & {\bf G292.2-00.5} & 0.80 & 0.73&YNG/YNG \\
  19$^*$ & 3FGL~J1124.5-5915 & 35.08 & {\bf G292.0+01.8/MSH11-54} & 1.0 & 0.99 & YNG/YNG \\ 
  20 &  3FGL~J1209.1-5224 & 4.83 &   {\bf G296.5+10.0}  & 0.00 & 0.01 & -/- \\ 
  21 &  3FGL~J1212.2-6251 & 8.47 &                            & 0.75 & 0.70 & YNG/YNG \\ 
  22 &  3FGL~J1214.0-6236 & 13.88 & {\bf G298.6--00.0}  & 0.94 & 0.81 & YNG/YNG \\ 
  23 &  3FGL~J1305.7-6241 & 11.12 & G304.6+00.1/Kes 17  & 0.14 & 0.26 & YNG/YNG \\ 
  24 &  3FGL~J1345.1-6224 & 10.53 &                            & 0.99 & 0.67 & YNG/YNG \\ 
  25 &  3FGL~J1441.5-5955c & 6.70 &  G316.3--00.0/MSH 14-57 & 0.68 & 0.39 & YNG/YNG \\ 
  26 &  3FGL~J1549.1-5347c & 11.59 &                            & 1.00 & 0.78 & YNG/YNG \\ 
  27 &  3FGL~J1551.1-5610 & 5.98 &  {\bf G326.3--01.8}  & 0.04 & 0.05 & -/-  \\ 
  28 &  3FGL~J1552.9-5610 & 18.36 & {\bf G326.3--01.8}  & 0.91 & 0.38 & YNG/YNG \\ 
  29 &  3FGL~J1615.3-5146e & 19.79 &                            & 0.82 & 0.30 & YNG/-  \\ 
  30 &  3FGL~J1628.9-4852 & 8.53 &                            & 1.00 & 0.85 & YNG/YNG \\ 
  31 &  3FGL~J1636.2-4709c & 10.45 &                            & 0.99 & 0.77 & YNG/YNG \\ 
  32 &  3FGL~J1636.2-4734 & 22.22 & SNR G337.0-00.1   & 1.00 & 0.83 & YNG/YNG \\ 
  33 &  3FGL~J1638.6-4654 & 13.39 & G337.8--00.1/Kes 41 & 0.96 & 0.77 & YNG/YNG \\ 
  34 &  3FGL~J1640.4-4634c & 10.41 &                            & 0.16 & 0.28 & YNG/YNG \\ 
  35 &  3FGL~J1641.1-4619c & 7.93 &                            & 0.77 & 0.57 & YNG/YNG \\ 
  36 &  3FGL~J1645.9-5420 & 4.98 &                            & 0.83 & 0.31 & MSP/MSP \\ 
  37 &  3FGL~J1713.5-3945e & 14.40 &  {\bf G347.3--00.5/RXJ1713.7-3946} & 0.34 & 0.25 & YNG/-  \\ 
  38 &  3FGL~J1714.5-3832 & 29.23 & {\bf G348.5+00.1/CTB37A} & 1.00 & 0.73 & YNG/YNG \\ 
  39 &  3FGL~J1718.0-3726 & 13.08 & {\bf G349.7+00.2} & 0.59 & 0.42 & YNG/YNG \\ 
  40 &  3FGL~J1722.9-4529 & 5.91 &                            & 0.23 & 0.32 & MSP/MSP \\ 
  41 &  3FGL~J1725.1-2832 & 6.15 &                            & 0.00 & 0.18 & YNG/YNG \\ 
  42 &  3FGL~J1728.0-4606 & 6.78 &                            & 0.23 & 0.29 & MSP/MSP \\ 
  43 &  3FGL~J1729.5-2824 & 6.73 &                            & 0.98 & 0.53 & YNG/YNG \\ 
  44 &  3FGL~J1737.3-3214c & 5.85 & G356.3--00.3  & 0.48 & 0.39 & YNG/YNG \\ 
  45 &  3FGL~J1741.1-3053 & 11.39 &  {\bf G357.7--00.1/MSH17-39} & 0.97 & 0.78 & YNG/YNG \\ 
  46 &  3FGL~J1745.1-3011 & 13.65 &                            & 1.00 & 0.77 & YNG/YNG \\ 
  47 &  3FGL~J1745.6-2859c & 11.97 &                            & 0.99 & 0.54 & YNG/YNG \\ 
  48 &  3FGL~J1746.3-2851c & 20.99 & PWN G0.13-0.11             & 1.00 & 0.85 & YNG/YNG \\ 
  49 &  3FGL~J1801.3-2326e & 62.71 & {\bf G006.4--00.1/W28} & 0.99 & 0.70 & YNG/-  \\ 
  50 &  3FGL~J1805.6-2136e & 33.18 & {\bf G008.7--00.1/W30}  & 0.91 & 0.51 & YNG/-  \\ 
  51 &  3FGL~J1810.1-1910 & 5.12 &                            & 0.03 & 0.35 & YNG/YNG \\ 
  52 &  3FGL~J1811.3-1927c & 4.39 &                            & 0.97 & 0.58 & YNG/YNG \\ 
  53 &  3FGL~J1817.2-1739 & 6.42 &                            & 0.49 & 0.60 & YNG/YNG \\ 
  54 &  3FGL~J1818.7-1528 & 13.25 &                            & 0.89 & 0.58 & YNG/YNG \\ 
  55 &  3FGL~J1828.4-1121 & 9.89 &  {\bf G020.0--00.2} & 1.00 & 0.54 & YNG/YNG \\ 
  56 &  3FGL~J1829.7-1304 & 6.50 & G018.9--01.1                           & 0.94 & 0.53 & YNG/YNG \\ 
  57 &  3FGL~J1833.9-0711 & 7.00 &  {\bf G024.7+00.6}                          & 0.08 & 0.41 & YNG/YNG \\ 
  58 &  3FGL~J1834.5-0841 & 13.91 & {\bf G023.3--00.3/W41}                       & 0.59 & 0.39 & YNG/YNG \\ 
  59 &  3FGL~J1834.6-0659 & 9.55 &                            & 0.84 & 0.43 & YNG/YNG \\ 
  60 &  3FGL~J1840.1-0412 & 9.44 &                            & 1.00 & 0.86 & YNG/YNG \\ 
  61 &  3FGL~J1849.4-0057 & 13.15 & 3C 391                     & 0.97 & 0.63 & YNG/YNG \\ 
  62 &  3FGL~J1855.9+0121e & 68.29 & {\bf G034.7--00.4/W44}                        & 1.00 & 0.73 & YNG/-  \\ 
  63 &  3FGL~J1910.9+0906 & 41.79 & {\bf G043.3--00.2/W49B}                      & 0.99 & 0.66 & YNG/YNG \\ 
  64 &  3FGL~J1915.9+1112 & 10.70 & {\bf G045.7--00.4}                           & 0.99 & 0.68 & YNG/YNG \\ 
  65 &  3FGL~J1923.2+1408e & 74.97 & {\bf G049.2--00.7/W51C}                       & 1.00 & 0.65 & YNG/-  \\ 
  66 &  3FGL~J1951.6+2926 & 6.18 &                            & 0.77 & 0.48 & YNG/MSP \\ 
  67 &  3FGL~J2014.4+3606 & 4.48 &  G073.9+00.9                          & 0.99 & 0.61 & YNG/YNG \\ 
  68 &  3FGL~J2022.2+3840 & 12.77 &                            & 1.00 & 0.72 & YNG/YNG \\ 
  69 &  3FGL~J2045.2+5026e & 31.54 & {\bf G089.0+04.7/HB21}                      & 1.00 & 0.79 & YNG/-  \\ 
  70 &  3FGL~J2051.0+3040e & 37.85 & {\bf G074.0-08.5/Cygnus Loop}                & 1.00 & 0.94 & YNG/-  \\ 
  71 &  3FGL~J2225.8+6045 & 4.09 &                            & 0.00 & 0.01 & -/-  \\ 
  72 &  3FGL~J2301.2+5853 & 6.08 & {\bf G109.1--01.0/CTB109}                 & 0.05 & 0.08 & -/-  \\ 
  73 &  3FGL~J2323.4+5849 & 36.66 &  {\bf G111.7--02.1/Cas A}                      & 0.97 & 0.64 & YNG/YNG \\ 
   \hline
\hline
\caption{LR and RF predictions for 3FGL sources with known SNR/PWN
  associations, including 27/30 likely GeV SNRs (in bold, in column 4) and 7/14 marginal candidates, as reported in the ``1st Fermi LAT Supernova Remnant
  Catalog''~\citep{SNRCat}. $^*$3FGL~J1119.1-6127 and
  3FGL~J1124.5-5915 are associated with PSRs J1119-6127 and J1124-5916
  in 3FGL, rather than with their respective SNRs.}
\label{snrpwn}
\end{longtable}

\begin{longtable}{rlrlrrll}
  \hline
 & Source\_Name & Signif & ASSOC1 & LR & RF & BLR\_PSR &RF\_PSR \\ 
& 3FGL &  &  & P & P & Pred & Pred \\ 
  \hline
1 & 3FGL~J0240.5+6113 & 196.61 & LS I+61 303                & 0.37 & 0.33 & YNG & YNG \\ 
  2 & 3FGL~J1018.9-5856 & 52.25 & 1FGL J1018.6-5856          & 1.00 & 0.77 & YNG & YNG \\ 
  3 & 3FGL~J1045.1-5941 & 33.09 & Eta Carinae                & 0.98 & 0.72 & YNG & YNG \\ 
  4 & 3FGL~J1826.2-1450 & 35.06 & LS 5039                    & 1.00 & 0.68 & YNG & YNG \\ 
   \hline
\hline
\caption{List of gamma-ray binaries in 3FGL, along with the RF and LR predictions.} 
\label{binaries}
\end{longtable}

\begin{longtable}{rlllllrr}
  \hline
 & Source\_Name & RA & DEC & GLON & GLAT & PSR\_Out & AGN\_Out \\ 
  \hline
1 & 3FGL~J0004.2+6757  &   1.055 &  67.9593 & 118.51 &   5.4940 & 76.78 & 218.93 \\ 
  2 & 3FGL~J0426.3+3510  &  66.588 &  35.1703 & 164.88 &  -9.6574 & 119.35 & 89.77 \\ 
  3 & 3FGL~J0534.5+2201s &  83.633 &  22.0199 & 184.55 &  -5.7812 & 131.12 & 83.51 \\ 
  4 & 3FGL~J1037.2-6052  & 159.319 & -60.8748 & 287.32 &  -2.1364 & 169.00 & 111.74 \\ 
  5 & 3FGL~J1151.8-6108  & 177.955 & -61.1438 & 295.80 &   0.8997 & 77.84 & 205.57 \\ 
  6 & 3FGL~J1325.2-5411  & 201.314 & -54.1859 & 307.92 &   8.3579 & 92.69 & 155.73 \\ 
  7 & 3FGL~J1624.2-3957  & 246.066 & -39.9566 & 341.10 &   6.6471 & 114.22 & 107.39 \\ 
  8 & 3FGL~J1632.8+3838  & 248.202 &  38.6477 &  61.74 &  42.8468 & 107.28 & 111.76 \\ 
  9 & 3FGL~J1729.9-0859  & 262.491 &  -8.9861 &  15.22 &  13.5760 & 134.00 & 96.12 \\ 
  10 & 3FGL~J1744.8-1557  & 266.215 & -15.9588 &  11.03 &   6.8759 & 131.85 & 78.40 \\ 
   \hline
\hline
\caption{List of the ten 3FGL sources with largest ($>$ 75) values of PSR and AGN outlyingness.} 
\label{out}
\end{longtable}

\begin{deluxetable}{lllllll}
\tablewidth{0pt}
\tablecaption{List of 3FGL sources for which the RF and LR classifiers
  are in agreement with each other but disagree with the 3FGL catalog classification.\label{misclassifications}}
\tablehead{
\colhead{3FGL Source} & \colhead{Signif.} & \colhead{ASSOC} & \colhead{CLASS}  & \colhead{Logistic\_Pred} &\colhead{RF\_Pred} & \colhead{YNG/MSP} \\
}
\startdata
3FGLJ0957.6+5523  &  112.9 & 4C +55.17 & fsrq & PSR (0.99) &PSR (0.23) & YNG/MSP\tablenotemark{a} \\
3FGLJ0217.3+6209  &  11.5 & TXS 0213+619 & bcu & PSR (0.99) & PSR (0.59) & YNG  \\
3FGLJ1136.1-7411  &  10.0 & PKS 1133-739 & bcu & PSR (0.67) & PSR (0.23)& MSP \\
3FGLJ1908.8-0130  &  9.0 & PKS 1133-739 & bcu & PSR (0.89) & PSR (0.28)& MSP \\
3FGLJ0744.8-4028  &  7.5 & NVSS J190836-012642 & bcu & PSR (0.70) & PSR (0.59)& MSP  \\
3FGLJ0401.4+2109 &  6.2 & PMN J0744-4032 & bcu & PSR (0.55) & PSR (0.50)& MSP  \\
3FGLJ0505.3-0422  &  5.2 & TXS 0358+210 & fsrq & PSR (0.27) & PSR (0.38)& MSP  \\
3FGLJ0333.4+4003 &  5.0 & S3 0503-04 & fsrq & PSR (0.46) &PSR (0.40) &MSP  \\
3FGLJ1513.1-1014  &  4.6 & B3 0330+399 & bcu & PSR (0.20) & PSR (0.33)& MSP  \\
3FGLJ1207.6-4537  & 4.6  & PKS 1511-100 & fsrq & PSR (0.72) & PSR (0.20)& MSP  \\
3FGLJ1757.1+1533 & 4.5 & PMN J1207-4531 & fsrq & PSR (0.13) & PSR (0.30)& MSP  \\
3FGLJ1136.1+7523 &  4.3 &  PSR J1137+7528 & PSR & AGN (0.00) &AGN (0.04)& --  \\
3FGLJ1525.2-5905  &  4.1 & PMN J1524-5903 & bcu & PSR (0.23) &PSR (0.52)& YNG  \\
\enddata
\tablenotetext{a}{RF predicts YNG, while Boosted LR predicts MSP.}
\end{deluxetable}

\begin{center}
\begin{landscape}
\begin{longtable}{cccccccc}
\hline \hline
\multirow{2}{*}{3FGL Name} & J2000 coord. & Counts rate & N$_\textrm{H}$ & $\Gamma_{X}$ & Flux$_{[0.3-10 \textrm{ keV}]}$ & \multirow{2}{*}{$\displaystyle\frac{\mbox{F$_{\gamma}$}}{\mbox{F$_X$}}$} & UL\\
  & RA Dec [$^{\circ}$] (stat. err.$^a$) & $10^{-3}$ cts/s & $10^{22}$ cm$^{-2}$ &  & $10^{-14}$\,erg cm$^{-2}$ s$^{-1}$ & & $10^{-14}$\,erg cm$^{-2}$ s$^{-1}$\\
\hline
\multirow{3}{*}{J0212.1+5320} & 33.0439, 53.3607 (0.02'') & 212.47$\pm$4.31 & 0.16$^{+0.03}_{-0.03}$ & 1.19$^{+0.06}_{-0.06}$ & 206.4$^{+7.0}_{-7.1}$ & 8$^{+1}_{-1}$ & \multirow{3}{*}{0.76}\\
 & 33.0587, 53.3223 (0.17'') & 1.05$\pm0.25$ & 0.16$^b$ & 2$^b$ & 1.78$^{+0.57}_{-0.53}$ & $963^{+525}_{-299}$\\
 & 33.0359, 53.3571 (0.37'') & 1.01$\pm$0.29 & 0.16$^b$ & 2$^b$ & 0.79$^{+0.23}_{-0.25}$ & $2170^{+1274}_{-641}$\\
\hline
\multirow{4}{*}{J0933.9--6232} & 143.5024, --62.5646 (0.07'') & 7.50$\pm$0.56 & 0.25 & 4.33$^{+0.22}_{-0.21}$ & 22.37$^{+1.98}_{-1.82}$ & 55$^{+10}_{-9}$ & \multirow{4}{*}{0.52}\\
 & 143.5049, --62.5077 (0.18'') & 1.55$\pm$0.26 & 0.25$^b$ & 2$^b$ & 0.54$^{+0.18}_{-0.17}$ & 2272$^{+1323}_{-730}$\\
 & 143.4916, --62.5239 (0.25'') & 1.18$\pm$0.22 & 0.25$^b$ & 3.04$^{+0.96}_{-0.97}$ & 1.33$^{+1.48}_{-0.25}$ & 922$^{+309}_{-504}$\\
 & 143.4357, --62.5548 (0.34'') & 0.70$\pm$0.18 & 0.25$^b$ & 2$^b$ & 1.22$^{+0.19}_{-0.24}$ & 1000$^{+352}_{-218}$\\
\hline
\multirow{1}{*}{J1035.7--6720} & 158.8652, --67.3371 (1.38'') & 5.70$\pm$0.96 & 0.2$^b$ & 2.91$_{-0.40}^{+0.46}$ & 3.06$_{-0.50}^{+0.97}$ & 848$^{+218}_{-240}$ & 0.90\\
\hline
\multirow{1}{*}{J1214.0--6236} & 183.4307, --62.5936 (0.29'') & 0.82$\pm$0.21 & 1.5$^b$ & 2$^b$ & 3.92$_{-0.90}^{+1.07}$ & 1131$_{-302}^{+345}$ & 4.5\\
\hline
\multirow{1}{*}{J1405.4--6119} & 211.3103, --61.3077 (0.19'') & 5.28$\pm$0.67 & 1.9$^b$ & 2$^b$ & 30.7$\pm$3.5 & 330$\pm$43 & 7.7\\
\hline
\multirow{1}{*}{J1744.1--7619} & 266.0030, --76.3205 (1.15'') & 5.36$\pm$0.72 & 0.08$^b$ & 2.71$_{-0.39}^{+0.40}$ & 1.92$_{-0.39}^{+0.59}$ & 1172$_{-327}^{+383}$ & 0.36\\
\hline\\
\multicolumn{8}{l}{\footnotesize{$^a$We report only the $1\sigma$
  statistical error. The $1\sigma$ systematic error is $\sim$1.5'' for {\em
  XMM-Newton} sources and $\sim$0.8'' for {\em Chandra} sources.}}\\
\multicolumn{8}{l}{\footnotesize{$^b$Due to the low statistics in these sources, we fixed this parameter in the spectral analysis.}}\\
\caption{Summary of the X-ray parameters of sources detected within
  the 95\% confidence error ellipse of the {\em Fermi}-LAT source, as
  discussed in the text. Here, we report the name of the 3FGL
  unassociated source, and for each X-ray plausible counterpart the
  best-fit position, the count rate, the best-fit column density and
  photon index, the unabsorbed X-ray flux in the 0.3--10 keV energy
  band, and the $\gamma$-to-X flux ratio. For each observation, we
  also report the minimum X-ray unabsorbed flux required for a
  5$\sigma$ detection of a source. All uncertainties are reported at the 68\% confidence level.}
\label{counterparts2}
\end{longtable}
\end{landscape}
\end{center}

\begin{longtable}{lrrrrrlrr}
  \hline
NAME & RA & DEC & ERR & RATE & FLUX & 3FGL\_NAME & R95 & SEP \\ 
  \hline
1FGL\_J0212.3+5319\_01 & 33.04 & 53.36 & 4.31 & 21.00 & 3.5E-13 &3FGLJ0212.1+5320 & 0.03 & 0.02 \\ 
3FGL\_J0225.8+6159\_01 & 36.51 & 61.97 & 5.64 & 4.20 & 5.6E-14 & 3FGLJ0225.8+6159 & 0.09 & 0.02 \\ 
1FGL\_J0523.5-2529\_01 & 80.82 & -25.46 & 4.67 & 4.40 & 8.9E-14 & 3FGLJ0523.3-2528 & 0.04 & 0.02 \\ 
1FGL\_J0523.5-2529\_02 & 80.85 & -25.46 & 5.14 & 2.80 & 5.7E-14 & 3FGLJ0523.3-2528 & 0.04 & 0.02 \\ 
2FGL\_J0539.3-0323\_01 & 84.88 & -3.57 & 6.12 & 4.20 & 6.8E-14 & 3FGLJ0538.8-0341 & 0.22 & 0.21 \\ 
1FGL\_J0737.4-3239\_02 & 114.31 & -32.61 & 5.76 & 5.00 & 7.5E-14  & 3FGLJ0737.2-3233 & 0.10 & 0.05 \\ 
1FGL\_J0737.4-3239\_01 & 114.41 & -32.55 & 5.49 & 4.90 & 7.3E-14 & 3FGLJ0737.2-3233 & 0.10 & 0.08 \\ 
2FGL\_J0802.7-5615\_03 & 120.60 & -56.09 & 5.89 & 1.80 & 3.0E-14 & 3FGLJ0802.3-5610 & 0.10 & 0.07 \\ 
3FGL\_J0826.3-5056\_01 & 126.60 & -50.96 & 5.82 & 8.00 & 1.2E-13 & 3FGLJ0826.3-5056 & 0.08 & 0.01 \\ 
1FGL\_J0838.6-2828\_01 & 129.68 & -28.45 & 3.55 & 105.00 & 1.8E-12 & 3FGLJ0838.8-2829 & 0.06 & 0.04 \\ 
1FGL\_J0838.6-2828\_05 & 129.74 & -28.52 & 5.11 & 0.80 & 1.4E-14 & 3FGLJ0838.8-2829 & 0.06 & 0.04 \\ 
3FGL\_J0855.4-4818\_01 & 133.77 & -48.25 & 5.35 & 8.00 & 1.1E-13 & 3FGLJ0855.4-4818 & 0.10 & 0.08 \\ 
3FGL\_J0859.3-4732\_02 & 134.78 & -47.51 & 3.78 & 19.00 & 2.6E-13 & 3FGLJ0859.3-4732 & 0.12 & 0.06 \\ 
3FGL\_J0859.3-4732\_05 & 134.78 & -47.55 & 4.18 & 1.80 & 2.5E-14 & 3FGLJ0859.3-4732 & 0.12 & 0.04 \\ 
3FGL\_J0859.3-4732\_03 & 134.79 & -47.57 & 4.09 & 3.40 & 4.7E-14 & 3FGLJ0859.3-4732 & 0.12 & 0.04 \\ 
3FGL\_J0859.3-4732\_07 & 134.99 & -47.55 & 4.74 & 1.30 & 1.8E-14 & 3FGLJ0859.3-4732 & 0.12 & 0.09 \\ 
3FGL\_J0859.3-4732\_01 & 135.02 & -47.56 & 3.84 & 16.00 & 2.2E-13 & 3FGLJ0859.3-4732 & 0.12 & 0.12 \\ 
3FGL\_J1043.6-5930\_14 & 160.82 & -59.56 & 3.90 & 1.30 & 1.6E-14 & 3FGLJ1043.6-5930 & 0.08 & 0.07 \\ 
1FGL\_J1045.2-5942\_23 & 160.92 & -59.58 & 3.53 & 0.79 & 9.5E-15 & 3FGLJ1043.6-5930 & 0.08 & 0.07 \\ 
3FGL\_J1043.6-5930\_13 & 160.93 & -59.58 & 3.58 & 2.00 & 2.4E-14 & 3FGLJ1043.6-5930 & 0.08 & 0.07 \\ 
3FGL\_J1043.6-5930\_02 & 160.99 & -59.55 & 3.53 & 43.30 & 5.2E-13 & 3FGLJ1043.6-5930 & 0.08 & 0.06 \\ 
1FGL\_J1045.2-5942\_05 & 160.99 & -59.55 & 3.51 & 16.70 & 2.0E-13 & 3FGLJ1043.6-5930 & 0.08 & 0.06 \\ 
3FGL\_J1047.3-6005\_01 & 161.79 & -60.07 & 4.76 & 34.00 & 4.0E-13 & 3FGLJ1047.3-6005 & 0.08 & 0.04 \\ 
3FGL\_J1050.6-6112\_01 & 162.67 & -61.27 & 5.89 & 9.00 & 1.1E-13 & 3FGLJ1050.6-6112 & 0.12 & 0.06 \\ 
1FGL\_J1106.2-1752\_02 & 166.58 & -17.76 & 6.93 & 3.30 & 6.3E-14 & 3FGLJ1106.6-1744 & 0.12 & 0.08 \\ 
1FGL\_J1106.2-1752\_01 & 166.66 & -17.79 & 6.52 & 3.40 & 6.5E-14 & 3FGLJ1106.6-1744 & 0.12 & 0.05 \\ 
1FGL\_J1119.9-2205\_02 & 170.00 & -22.08 & 3.92 & 2.30 & 4.4E-14 & 3FGLJ1119.9-2204 & 0.04 & 0.02 \\ 
3FGL\_J1120.6+0713\_02 & 170.18 & 7.22 & 4.83 & 5.90 & 1.1E-13 & 3FGLJ1120.6+0713 & 0.07 & 0.01 \\ 
3FGL\_J1126.8-5001\_01 & 171.60 & -50.14 & 4.12 & 27.00 & 4.7E-13 & 3FGLJ1126.8-5001 & 0.16 & 0.14 \\ 
3FGL\_J1126.8-5001\_02 & 171.81 & -49.90 & 5.89 & 3.50 & 6.1E-14 & 3FGLJ1126.8-5001 & 0.16 & 0.14 \\ 
3FGL\_J1151.8-6108\_01 & 177.93 & -61.16 & 5.89 & 7.00 & 9.2E-14 & 3FGLJ1151.8-6108 & 0.10 & 0.02 \\ 
1FGL\_J1232.2-5118\_02 & 187.87 & -51.16 & 5.49 & 2.40 & 4.0E-14 & 3FGLJ1231.6-5113 & 0.11 & 0.06 \\ 
3FGL\_J1311.8-6230\_11 & 197.92 & -62.55 & 3.72 & 1.10 & 1.3E-14 & 3FGLJ1311.8-6230 & 0.07 & 0.05 \\ 
2FGL\_J1329.7-6108\_02 & 202.41 & -61.12 & 5.96 & 4.10 & 5.4E-14 & 3FGLJ1329.8-6109 & 0.05 & 0.05 \\ 
3FGL\_J1405.4-6119\_02 & 211.31 & -61.31 & 5.05 & 3.80 & 3.9E-14 & 3FGLJ1405.4-6119 & 0.03 & 0.02 \\ 
1FGL\_J1405.1-6123c\_02 & 211.31 & -61.31 & 4.90 & 3.20 & 3.3E-14 & 3FGLJ1405.4-6119 & 0.03 & 0.02 \\ 
3FGL\_J1421.0-2431\_01 & 215.15 & -24.51 & 5.24 & 8.00 & 1.5E-13 & 3FGLJ1421.0-2431 & 0.12 & 0.10 \\ 
3FGL\_J1445.7-5925\_01 & 221.43 & -59.53 & 5.76 & 17.00 & 1.9E-13 & 3FGLJ1445.7-5925 & 0.15 & 0.11 \\ 
3FGL\_J1544.1-2555\_01 & 235.93 & -25.94 & 6.93 & 9.00 & 1.5E-13 & 3FGLJ1544.1-2555 & 0.13 & 0.09 \\ 
3FGL\_J1544.6-1125\_08 & 236.15 & -11.50 & 4.20 & 0.90 & 1.5E-14 & 3FGLJ1544.6-1125 & 0.08 & 0.07 \\ 
3FGL\_J1544.6-1125\_01 & 236.16 & -11.47 & 3.55 & 71.00 & 1.2E-12 & 3FGLJ1544.6-1125 & 0.08 & 0.04 \\ 
3FGL\_J1544.6-1125\_05 & 236.17 & -11.42 & 4.17 & 1.40 & 2.4E-14& 3FGLJ1544.6-1125 & 0.08 & 0.01 \\ 
2FGL\_J1617.3-5336\_01 & 244.19 & -53.73 & 5.20 & 9.00 & 1.2E-13 & 3FGLJ1616.8-5343 & 0.09 & 0.01 \\ 
1FGL\_J1624.0-4041\_05 & 246.04 & -40.70 & 4.13 & 0.60 & 9.4E-15 & 3FGLJ1624.2-4041 & 0.04 & 0.02 \\ 
3FGL\_J1624.1-4700\_01 & 246.13 & -46.97 & 6.78 & 29.00 & 3.8E-13 & 3FGLJ1624.1-4700 & 0.09 & 0.07 \\ 
1FGL\_J1625.8-2429c\_04 & 246.53 & -24.46 & 6.41 & 6.00 & 9.8E-14 & 3FGLJ1626.2-2428c & 0.06 & 0.04 \\ 
2FGL\_J1626.4-4408\_01 & 246.59 & -44.12 & 5.82 & 6.00 & 9.0E-14 & 3FGLJ1626.3-4406 & 0.13 & 0.01 \\ 
1FGL\_J1627.6+3218\_01 & 246.93 & 32.35 & 6.65 & 9.00 & 1.8E-13 & 3FGLJ1627.8+3217 & 0.07 & 0.06 \\ 
3FGL\_J1653.6-0158\_06 & 253.41 & -1.98 & 4.33 & 3.00 & 5.3E-14 & 3FGLJ1653.6-0158 & 0.04 & 0.01 \\ 
3FGL\_J1703.9-4843\_12 & 255.83 & -48.67 & 3.83 & 0.60 & 9.1E-15 & 3FGLJ1703.9-4843 & 0.13 & 0.12 \\ 
3FGL\_J1703.9-4843\_13 & 255.88 & -48.69 & 3.89 & 0.60 & 9.1E-15 & 3FGLJ1703.9-4843 & 0.13 & 0.08 \\ 
3FGL\_J1703.9-4843\_11 & 255.90 & -48.79 & 3.83 & 0.70 & 1.1E-14 & 3FGLJ1703.9-4843 & 0.13 & 0.09 \\ 
3FGL\_J1703.9-4843\_03 & 255.96 & -48.76 & 3.82 & 1.20 & 1.8E-14 & 3FGLJ1703.9-4843 & 0.13 & 0.04 \\ 
3FGL\_J1710.6-4317\_01 & 257.78 & -43.40 & 4.10 & 15.00 & 2.1E-13 & 3FGLJ1710.6-4317 & 0.15 & 0.14 \\ 
3FGL\_J1734.7-2930\_01 & 263.79 & -29.50 & 6.41 & 12.00 & 1.7E-13 & 3FGLJ1734.7-2930 & 0.10 & 0.08 \\ 
3FGL\_J1749.2-2911\_13 & 267.34 & -29.25 & 3.67 & 0.90 & 1.2E-14 & 3FGLJ1749.2-2911 & 0.08 & 0.06 \\ 
3FGL\_J1749.2-2911\_11 & 267.37 & -29.26 & 3.68 & 1.10 & 1.4E-14  & 3FGLJ1749.2-2911 & 0.08 & 0.08 \\ 
3FGL\_J1754.0-2930\_01 & 268.50 & -29.49 & 4.27 & 20.00 & 2.9E-13 & 3FGLJ1754.0-2930 & 0.11 & 0.02 \\ 
3FGL\_J1808.4-3703\_04 & 272.01 & -37.06 & 3.57 & 1.79 & 3.1E-14 & 3FGLJ1808.4-3703 & 0.13 & 0.09 \\ 
3FGL\_J1808.4-3703\_19 & 272.01 & -36.97 & 3.57 & 0.52 & 9.0E-15 & 3FGLJ1808.4-3703 & 0.13 & 0.12 \\ 
3FGL\_J1808.4-3703\_33 & 272.03 & -37.15 & 3.60 & 0.28 & 4.8E-15 & 3FGLJ1808.4-3703 & 0.13 & 0.12 \\ 
3FGL\_J1808.4-3703\_14 & 272.05 & -37.03 & 3.57 & 0.59 & 1.0E-14 & 3FGLJ1808.4-3703 & 0.13 & 0.06 \\ 
2FGL\_J1808.3-3356\_03 & 272.05 & -33.96 & 4.49 & 2.10 & 3.6E-14 & 3FGLJ1808.3-3357 & 0.09 & 0.04 \\ 
3FGL\_J1808.4-3703\_32 & 272.10 & -37.07 & 3.57 & 0.41 & 7.1E-15 & 3FGLJ1808.4-3703 & 0.13 & 0.03 \\ 
2FGL\_J1808.3-3356\_01 & 272.10 & -33.94 & 4.57 & 2.20 & 3.8E-14 & 3FGLJ1808.3-3357 & 0.09 & 0.03 \\ 
3FGL\_J1808.4-3703\_38 & 272.10 & -36.99 & 3.50 & 0.39 & 6.7E-15 & 3FGLJ1808.4-3703 & 0.13 & 0.06 \\ 
3FGL\_J1808.4-3703\_01 & 272.11 & -36.98 & 3.50 & 296.60 & 5.1E-12 & 3FGLJ1808.4-3703 & 0.13 & 0.07 \\ 
3FGL\_J1808.4-3703\_40 & 272.12 & -36.93 & 3.55 & 0.35 & 6.0E-15 & 3FGLJ1808.4-3703 & 0.13 & 0.12 \\ 
3FGL\_J1808.4-3703\_28 & 272.15 & -37.04 & 3.57 & 0.42 & 7.2E-15 & 3FGLJ1808.4-3703 & 0.13 & 0.03 \\ 
3FGL\_J1808.4-3703\_13 & 272.16 & -36.98 & 3.54 & 0.69 & 1.2E-14 & 3FGLJ1808.4-3703 & 0.13 & 0.08 \\ 
3FGL\_J1808.4-3703\_09 & 272.18 & -37.04 & 3.57 & 1.04 & 1.8E-14 & 3FGLJ1808.4-3703 & 0.13 & 0.04 \\ 
3FGL\_J1808.4-3703\_17 & 272.19 & -37.09 & 3.58 & 0.50 & 8.6E-15 & 3FGLJ1808.4-3703 & 0.13 & 0.07 \\ 
3FGL\_J1808.4-3703\_25 & 272.21 & -37.08 & 3.58 & 0.47 & 8.1E-15 & 3FGLJ1808.4-3703 & 0.13 & 0.08 \\ 
3FGL\_J1808.4-3703\_11 & 272.23 & -37.11 & 3.61 & 0.60 & 1.0E-14 & 3FGLJ1808.4-3703 & 0.13 & 0.10 \\ 
3FGL\_J1808.4-3703\_36 & 272.25 & -37.05 & 3.60 & 0.31 & 5.3E-15 & 3FGLJ1808.4-3703 & 0.13 & 0.10 \\ 
3FGL\_J1808.4-3703\_02 & 272.26 & -37.00 & 3.56 & 2.90 & 5.0E-14 & 3FGLJ1808.4-3703 & 0.13 & 0.12 \\ 
3FGL\_J1808.4-3703\_41 & 272.27 & -37.03 & 3.59 & 0.27 & 4.7E-15 & 3FGLJ1808.4-3703 & 0.13 & 0.12 \\ 
2FGL\_J1819.3-1523\_01 & 274.70 & -15.46 & 4.95 & 8.00 & 9.7E-14 & 3FGLJ1818.7-1528 & 0.09 & 0.01 \\ 
2FGL\_J1828.7+3231\_01 & 277.31 & 32.58 & 4.51 & 14.00 & 2.6E-13 & 3FGLJ1829.2+3229 & 0.15 & 0.08 \\ 
2FGL\_J1828.7+3231\_02 & 277.40 & 32.50 & 6.41 & 2.80 & 5.1E-14 & 3FGLJ1829.2+3229 & 0.15 & 0.08 \\ 
3FGL\_J1844.3-0344\_09 & 281.07 & -3.72 & 3.96 & 0.44 & 4.8E-15 & 3FGLJ1844.3-0344 & 0.06 & 0.04 \\ 
3FGL\_J1844.3-0344\_03 & 281.14 & -3.78 & 3.84 & 1.00 & 1.1E-14 & 3FGLJ1844.3-0344 & 0.06 & 0.05 \\ 
2FGL\_J1908.8-0132\_01 & 287.18 & -1.50 & 5.02 & 5.20 & 8.0E-14& 3FGLJ1908.8-0130 & 0.08 & 0.03 \\ 
1FGL\_J1908.5-0138\_01 & 287.18 & -1.50 & 5.00 & 4.50 & 7.0E-14 & 3FGLJ1908.8-0130 & 0.08 & 0.03 \\ 
3FGL\_J1921.6+1934\_01 & 290.31 & 19.67 & 5.02 & 16.00 & 2.2E-13 & 3FGLJ1921.6+1934 & 0.13 & 0.13 \\ 
3FGL\_J1946.4-5403\_02 & 296.64 & -54.04 & 4.71 & 4.40 & 8.5E-14 & 3FGLJ1946.4-5403 & 0.06 & 0.02 \\ 
2FGL\_J2009.2-1505\_01 & 302.16 & -15.08 & 6.31 & 4.00 & 7.2E-14 & 3FGLJ2009.2-1458 & 0.53 & 0.17 \\ 
2FGL\_J2133.5-6431\_01 & 323.30 & -64.64 & 4.13 & 13.00 & 2.5E-13 & 3FGLJ2133.0-6433 & 0.10 & 0.09 \\ 
2FGL\_J2133.5-6431\_02 & 323.35 & -64.58 & 5.31 & 1.80 & 3.5E-14 & 3FGLJ2133.0-6433 & 0.10 & 0.04 \\ 
3FGL\_J2215.5+6122\_01 & 333.65 & 61.45 & 4.64 & 6.00 & 8.3E-14 & 3FGLJ2215.5+6122 & 0.14 & 0.13 \\ 
3FGL\_J2250.6+3308\_02 & 342.69 & 33.09 & 6.12 & 6.00 & 1.1E-13 & 3FGLJ2250.6+3308 & 0.13 & 0.06 \\ 
   \hline
\hline
\caption{List of plausible Swift X-ray counterparts to 3FGL sources
  predicted by both RF and LR to be pulsars. RA and DEC of X-ray
  source, in deg. ERR is the uncertainty in the X-ray position, in arcsec. RATE (counts ksec$^{-1}$),  FLUX is the
  0.1--2.4 keV flux (erg cm$^{-2}$ s$^{-1}$, see text for details). NB: All X-ray parameters taken from {\tt
    http://www.swift.psu.edu/unassociated/}.
  R95 is the 95\% uncertainty in the semi-major axis of the 3FGL LAT
  position. SEP is the angular separation between the gamma-ray and
  X-ray positions, in degrees.} 
\label{counterparts}
\end{longtable}

\rotate
\newpage

\newpage

\newpage

\newpage

\newpage

\newpage
\begin{figure}
\epsscale{1.0}
\includegraphics[width=6in]{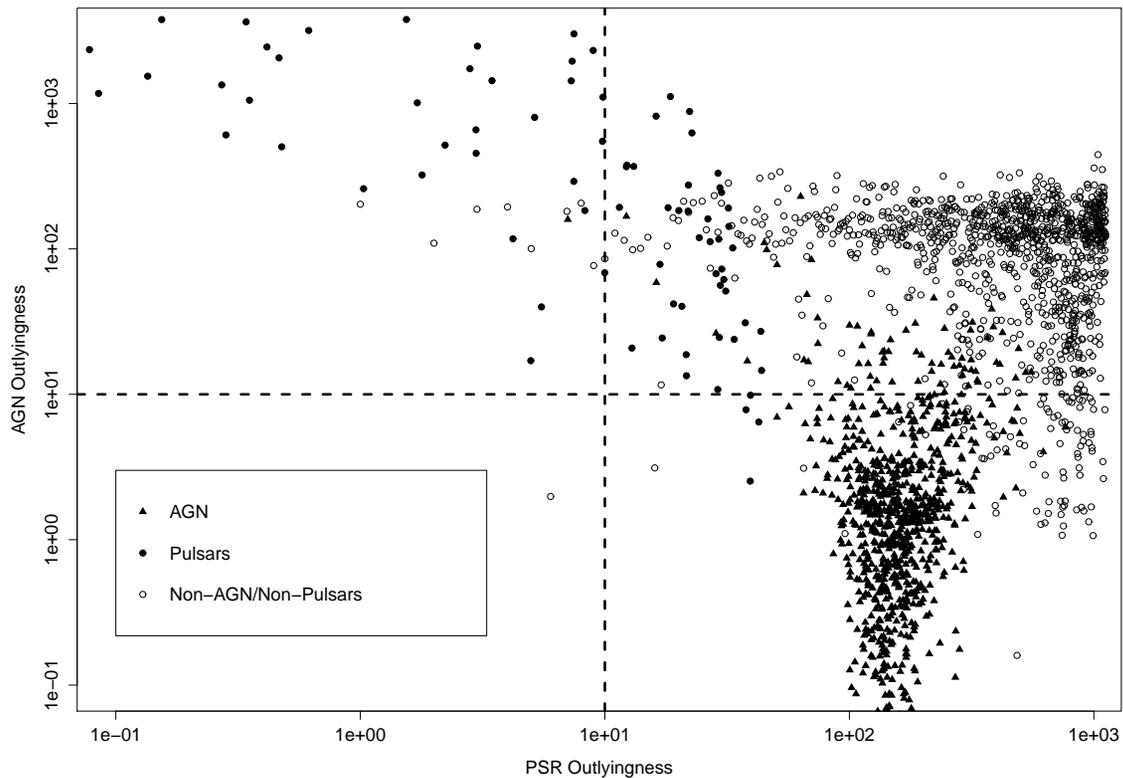}
\caption{``Outlyingness'' of all 3FGL sources, with respect to the PSR and
  AGN classes. The different symbols represent 3FGL sources associated with AGN (triangles),
  Pulsars (filled circles), and neither (empty circles). A large value ($>$10) along one axis implies the source is unlikely to
  belong to that class. A large value of ``outlyingness'' along both
  axes could imply a different gamma-ray source class altogether
  (i.e. non-Pulsar and non-AGN). Roughly 30 sources have
  ``outlyingness'' values above 50 in both axes.\label{outlyingness}}
\end{figure}

\newpage
\begin{figure}
\epsscale{1.0}
\includegraphics[width=6in]{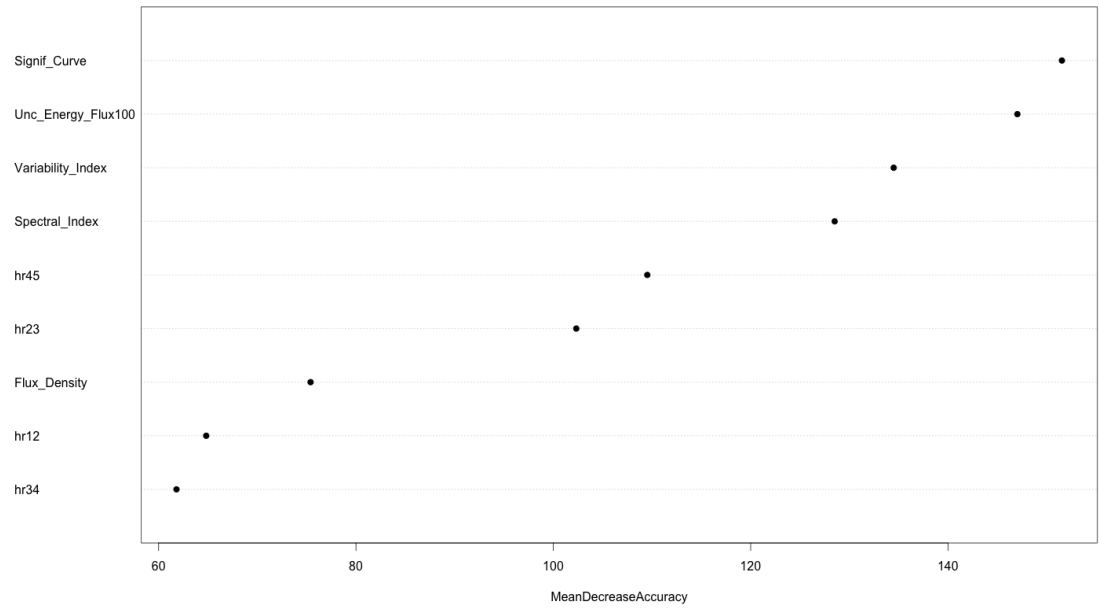}
\caption{Mean Decrease in Accuracy for the best Random Forest model to
  classify sources into AGN vs Pulsar.
\label{mda1}}
\end{figure}

\begin{figure}
\epsscale{1.0}
\includegraphics[width=6in]{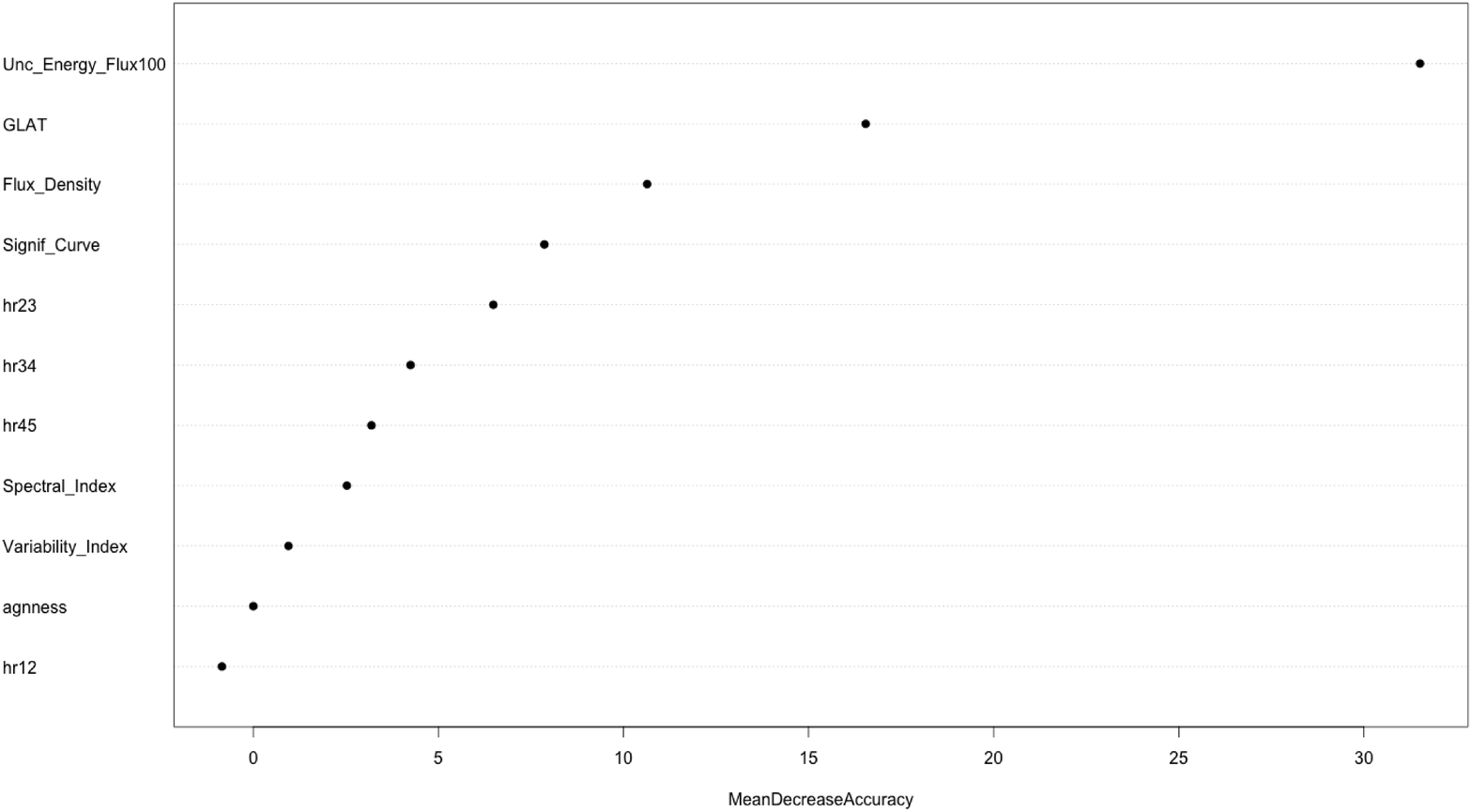}
\caption{Mean Decrease in Accuracy for the best Random Forest model to
  classify pulsars into `Young' (YNG) and  MSPs.
\label{mda2}}
\end{figure}

\newpage
\begin{figure}
\epsscale{1.0}
\includegraphics[width=5in]{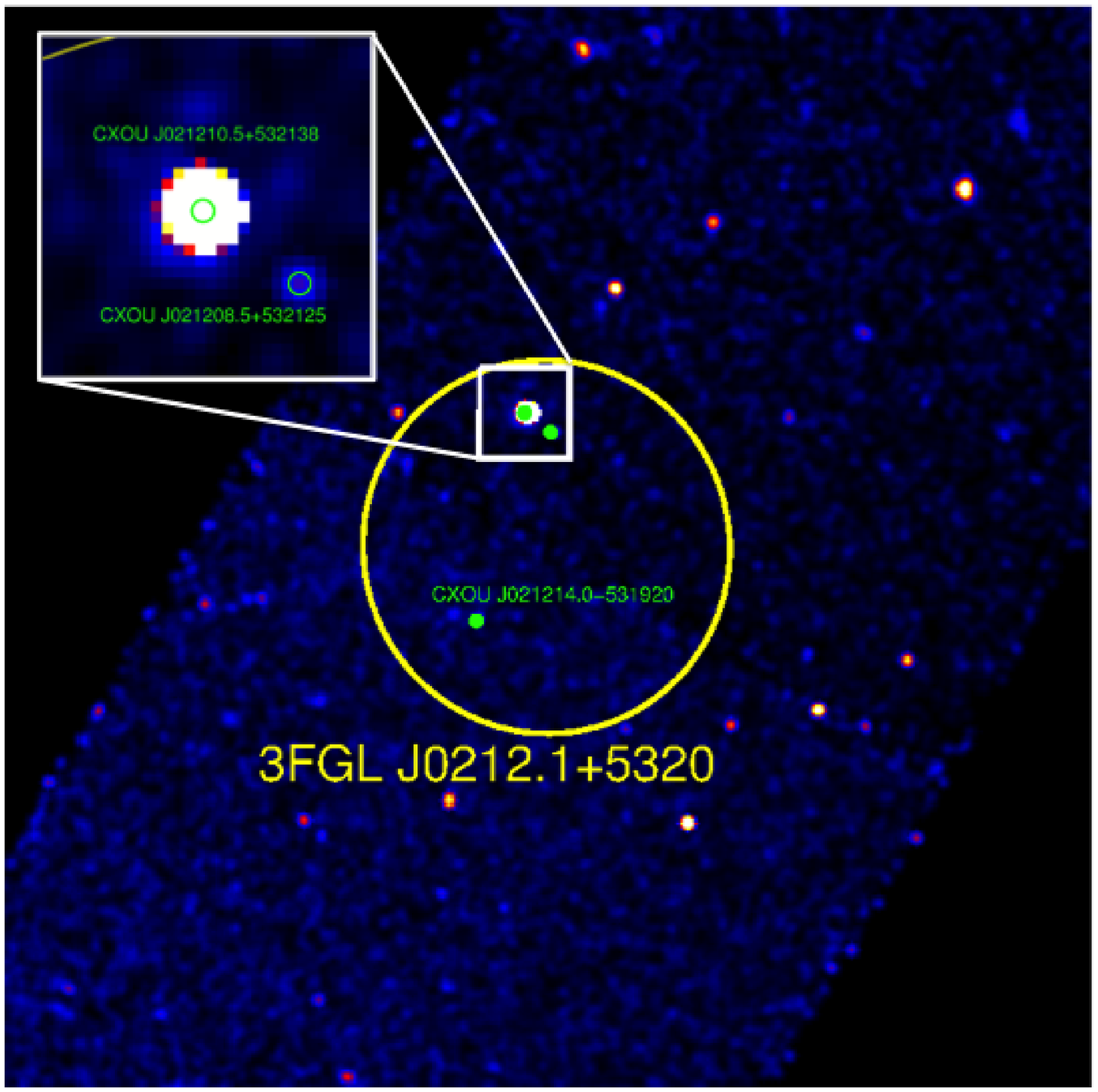}
\caption{{\it Chandra} 30 ks observation of 3FGL~J0212.1+5320,
  predicted by our algorithm to be a millisecond pulsar (MSP, see
  Table~\ref{unassoc}). The smoothed 0.3-10 keV ACIS-I
  exposure-corrected (10'x10') image includes the 95\% confidence LAT
  error ellipse (shown in yellow). The most probable X-ray
  counterpart (CXOU J021210.5+532138) is highlighted in a zoomed 1'x1' region, with a 2" radius
  source extraction region shown in green.
 \label{J0212}}
\end{figure}

\newpage
\begin{figure}
\epsscale{1.0}
\includegraphics[width=5in]{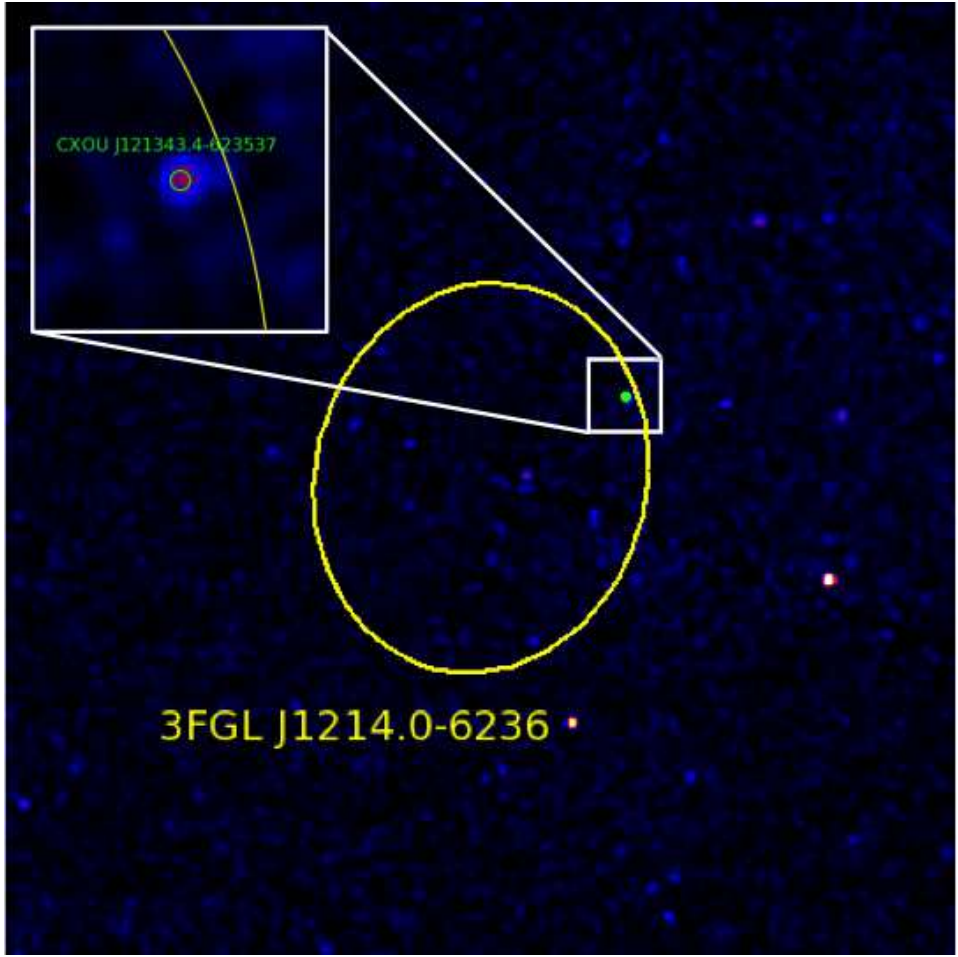}
\caption{{\it Chandra} 20 ks observation of 3FGL~J1214.0-6236,
  coincident with SNR G298.6--0.0, predicted by our algorithm to be a
  young pulsar (YNG, see Table~\ref{snrpwn}). The smoothed 0.3-10 keV ACIS-I
  exposure-corrected (13'x13') image includes the 95\% confidence LAT
  error ellipse (shown in yellow). The most probable X-ray
  counterpart (CXOU J121343.4-623537) is highlighted in a zoomed 1'x1' region, with a 2" radius
  source extraction region shown in green.
 \label{J1214}}
\end{figure}

\end{document}